%% file: revision1.tex
\RequirePackage{tikz}
\documentclass[smallextended]{svjour3}

\usepackage[margin=1in]{geometry}
\usepackage[toc]{appendix}
\usepackage[backend=biber, style=apa, sortcites=false, uniquelist=false,natbib=true,doi=false,isbn=false,url=false,eprint=false,pagetracker, ibidtracker=constrict]{biblatex}
\DeclareSourcemap{
  \maps{
    \map{
      \step[fieldset=entrysubtype, null]
      \step[fieldset=address, null]
      \step[fieldset=location, null]
    }
  }
}

\usepackage[colorlinks=true,citecolor=blue]{hyperref}
\usepackage{latexsym}
\usepackage{graphicx}

\usepackage{tikz}
\usepackage{pgfplots}
\usetikzlibrary{shapes,positioning,calc}

\usepackage{longtable}
\usepackage{multirow}
\usepackage{multicol}

\usepackage{amsmath}
\usepackage{amssymb}
\usepackage{dsfont}
\usepackage{textcomp}
\usepackage{textgreek}
\usepackage[french,english]{babel}
\usepackage{latexsym}
\usepackage{url}  
\usepackage{graphicx}  
\usepackage{url}
\usepackage{booktabs}
\usepackage{array}

\usepackage[acronym,nomain,nonumberlist]{glossaries}
\makeglossaries

\input{acronyms_en}

\title{How research programs come apart: the example of supersymmetry and the disunity of physics}

\author{Lucas Gautheron \\
  \texttt{lucas.gautheron@gmail.com} \\
  Elisa Omodei
}

\journalname{TBD}

\newcommand*{\affaddr}[1]{#1} 
\newcommand*{\affmark}[1][*]{\textsuperscript{#1}}

\author{Lucas Gautheron \protect\affmark[1,2]  \and
    Elisa Omodei \protect\affmark[3]}
    
\institute{
 Lucas Gautheron \\   \email{lucas.gautheron@gmail.com}\\ORCID: 0000-0002-3776-3373\\\\
 Elisa Omodei \\   \email{omodeie@ceu.edu}\\ORCID: 0000-0002-6748-5124\\\\
 \affaddr{\affmark[1]Interdisciplinary Centre for Science and Technology Studies (IZWT), University of Wuppertal, Germany}\\
 \affaddr{\affmark[2]Sciences Po, médialab, Paris, France}\\
 \affaddr{\affmark[3]Department of Network and Data Science, Central European University, Vienna, Austria}
}

\date{}

\begin{document}

\maketitle


\abstract{
According to Peter Galison, the coordination of different ``subcultures'' within a scientific field happens through local exchanges within ``trading zones''. In his view, the workability of such trading zones is not guaranteed, and science is not necessarily driven towards further integration. In this paper, we develop and apply quantitative methods (using semantic, authorship, and citation data from scientific literature), inspired by Galison's framework, to the case of the disunity of high-energy physics. We give prominence to supersymmetry, a concept that has given rise to several major but distinct research programs in the field, such as the formulation of a consistent theory of quantum gravity or the search for new particles. We show that ``theory'' and `phenomenology'' in high-energy physics should be regarded as distinct theoretical subcultures, between which supersymmetry has helped sustain scientific ``trades''. However, as we demonstrate using a topic model, the phenomenological component of supersymmetry research has lost traction and the ability of supersymmetry to tie these subcultures together is now compromised. Our work supports that even fields with an initially strong sentiment of unity may eventually generate diverging research programs and demonstrates the fruitfulness of the notion of trading zones for informing quantitative approaches to scientific pluralism.
}

\keywords{scientific pluralism; trading zones; topic models; citation networks; high-energy physics}


\section{Introduction}
\label{section:introduction}

This paper focuses on \gls{hep}, the field of physics concerned with the fundamental entities of nature, and ``supersymmetry'', a symmetry between the two basic types of particles in nature. The idea of supersymmetry has brought together many of the most significant developments in the field throughout the past 50 years, all the way from the highly abstract world of string theorists, deep down to the machinery of under-ground particle colliders. However, none of the discoveries that supersymmetry promised have materialized as expected; as much as supersymmetry may be necessary to theorists seeking to unify the forces of nature into a coherent picture, it is increasingly plausible that it will not be of much use to the experimentalists looking to find new particles. Throughout this case study, therefore, our work exhibits the disunity of science, by demonstrating that even scientific fields that have been strongly committed to unity, such as \gls{hep}, can eventually fail to coordinate various research efforts. Our paper is guided by the idea that empirical case-studies, although seemingly narrow in scope, do enrich our understanding of the nature of scientific enterprise (in this case, the nature of the coordination of diverse scientific cultures), and that quantitative studies of science should provide conceptually informed tools for carrying out such case-studies, preferably in ways that can be generalized for a variety of contexts. 

We start by presenting Galison's notions of subcultures and trading zones which is the framework for studying the plurality of science and the dynamics of interactions between scientific fields that underlies our investigation 
(\ref{section:galison}). We will then provide the necessary background knowledge for understanding the context of our case-study before laying out our hypotheses: i) that theory and phenomenology, over the historical period considered (1980--2020), are to be regarded as two distinct theoretical subcultures within high-energy physics; ii) that supersymmetry generated diverse research programs, some being phenomenological and some being more theoretical; and iii), that supersymmetry significantly contributed to sustain successful trades between theory and phenomenology until it was put in doubt by experimental data (\ref{section:hep_susy}). We then elaborate our motivation for addressing these hypotheses through quantitative methods (\ref{section:quantitative}). Then, section \ref{section:method} details the quantitative methods that were deployed in order to address each of the three claims put forward in the introduction. It starts with a description of the data on which our analysis rests and how it was collected (\ref{section:data}). Subsection \ref{section:method_subcultures} elaborates quantitative methods for assessing the level of semantic and social autonomy of certain categories (subcultures), and applies these methods to the two theoretical subcultures in \gls{hep}. The next subsection (\ref{section:method_plurality}) elaborates a methodology based on topic models in order to address the ``plasticity'' and ``plurality'' of supersymmetry, which can in principle be applied to all ``boundary objects'', i.e. those objects that can be traded between distinct subcultures while preserving and sustaining their distinctness. Finally, subsection \ref{section:method_trading_zone} provides a quantitative model for locating ``trading zones'' or more broadly concepts that enhance trades between subcultures (or scientific disciplines in general), and applies the model to the exchanges between the theoretical subcultures of \gls{hep}. Section \ref{section:application} reveals and interprets the results of these analyses. Finally, Section \ref{section:discussion} explores the consequences of this work, both for our case study (supersymmetry within \gls{hep}) and for the more general question of the plurality of science from a quantitative perspective.

\section{\label{section:background}Background}

\subsection{Subcultures and trading zones: Galison's approach to the plurality of science}\label{section:galison}

If science is a unified enterprise, what is the nature of the relationship between fields as diverse as physics, biology, psychology, or economics? Can we translate all the concepts of these disciplines into a basic (say, physical) scientific language, as Carnap proposed? Or, are all these fields so incommensurable and autonomous that it is impossible to translate their respective entities, laws, and explanations from one's language to another's, as proponents of a pluralistic view defend (e.g. \citet{Suppes1978,Dupre1983,Cartwright1999})? Disciplines themselves can be so diverse, too, that the nature of what makes their own unity is not necessarily obvious. For instance, the nature of the unity of physics has been the matter of much debate, with sometimes serious political implications: reductionist views (which imply that high-energy physics is the most fundamental, since it supposedly entail any higher-level theory) were mobilized to justify the funding of large particle physics facilities \citep{Cat1998}, potentially to the detriment of more ``useful'' projects, as certain condensed matter physicists argued \citep[Ch.~9]{martin2018solid}. Instead, the latter argued that macroscopic systems have emergent properties that cannot be derived from ``fundamental'' laws. They were most often proponents of a ``methodological'' form of unity (Ibid., p.~233), according to which the field is bound together by shared norms and conceptual tools \citep[p.~267]{Cat1998}, rather than by relations of logical deduction from the most fundamental to the least fundamental theories. This view provided an intellectual and philosophical basis for elevating the prestige of condensed matter physics \citep[p.~148--149]{martin2018solid}, thus putting condensed matter and high-energy physicists on a more equal footing.

Even within the subfield of particle physics, there is a strong contrast between theorists and experimentalists. In fact, the nature of the relationship between the objects manipulated by, say, experimentalists (for instance, tracks within a cloud-chamber, or electric signals from a sensor) and the more abstract entities manipulated by theorists (e.g. ``quarks'', ``gluons'', ``strings'', etc.) has been the subject of much philosophical debate. Inheriting a positivist view, some would grant experiment a more fundamental status, by defending its ability to provide robust empirical statements that could dictate theoretical  change. Others, such as Kuhn, argued that empirical statements cannot be isolated from a theoretical paradigm and emphasized the``primacy'' of theory \citep{Galison1988}\footnote{For instance, in his historical account of the discovery of quarks, \citet[p.~411]{pickering1984constructing} endorses the Kuhnian view: ``To attempt to choose between old- and new-physics [gauge] theories on the basis of a common set of phenomena [experimental facts] was impossible: the theories were integral parts of different worlds, and they were incommensurable''. Instead, Galison emphasizes the relative continuity and robustness of experimental ``facts'', across theoretical changes.}. It is in order to overcome this debate about the relationship between experiment and theory within the context of physics that Galison originally developed his concepts of subcultures and trading zones  \citep{galison1987how,galison1997image}. However, these notions may apply more generally whenever distinct scientific communities attempt to overcome difficulties to communicate and achieve coordination \citep[p.~8]{Collins2010}. Consequently it is useful in a much broader range of contexts than the narrow case of physics; for instance, it is generally useful for studying the dynamics of interactions between disciplines in science\footnote{For example, \citet{Kemman2021} describes Digital History as a trading zone.}. Below, we propose a brief summary of the concepts of subcultures and trading zones and the rationale for their introduction.

The notion of subcultures was introduced by \citet{galison1987how,Galison1988} in order to account for two characteristics of high-energy physics: first, that it is subject to a strong division of labor, such that ``theory'', ``experiment'' and ``instrumentation'' are carried out by different groups of people \citep[p.~138]{galison1987how}, with their own skill sets and bodies of knowledge; and second, that each of these ``subcultures'' are partially autonomous, i.e., none of them are completely subordinate to the others. We can highlight two tangible components of such subcultures: a social component--the community of practitioners--and a linguistic component--the language specific to each community.

For Galison, then, the question is what makes these subcultures part of a ``larger culture'' (physics), 
while retaining that their successful coordination is a ``contingent matter'' \citep[p.~18]{galison1997image}; and his answer is ``trading zones''. Trading zones allow knowledge to be exchanged across different subcultures, inasmuch as the practitioners of distinct communities can locally agree on the usefulness of certain constructs despite the distinctiveness of their respective languages, commitments, aims, and methodologies. That trading occurs within ``zones'' captures the fact that the exchange procedure is  ``local'' rather than ``global'', such that subcultures working out trades with each other can retain much of their autonomy in the process.

What kinds of goods may be subject to these ``trades''? Examples of trade-able goods are ``boundary objects'', i.e. ``objects that are both plastic enough to adapt to local needs and constraints of the several parties employing them, yet robust enough to maintain a common identity across sites'' \citep[p.~393]{Star1989}\footnote{In the context of physics, Darrigol's theoretical modules \citep[p.~214]{Darrigol2007}, or multi-purpose scientific instruments \citep[pp.~179--182]{Shinn2005}, may be other examples of such trade-able goods.}. Trading zones may give rise to a purposefully crafted inter-language that allows for further communication and coordination (a ``pidgin''). If the inter-language grows, it may turn into a full-blown language (a ``creole''); this signals the emergence and stabilization of a new scientific discipline of its own. 

Arguably, this is the process through which ``phenomenology'' -- a subfield of \gls{hep} at the boundary between theory and experiment -- has developed \citep[p.~837]{galison1997image}. However, we may wonder whether phenomenology is still merely dedicated to bridging the gap between the theoretical and experimental cultures, or whether it acquired enough autonomy to depart from the supremacy of abstract theory -- e.g., by relying on independent sources of inspiration for its own enterprise rather than by seeking to establish connections between high theory and experiment. In the following subsection we will suggest treating ``theory'' and ``phenomenology'' in high-energy physics as two distinct subcultures, such that they may both enjoy considerable autonomy and eventually fail to coordinate their developments -- thus extending the distinction made by Galison between theory, experiment and instrumentation.

\subsection{Supersymmetry across theory and phenomenology}
\label{section:hep_susy}

\subsubsection{Theory and phenomenology as distinct subcultures within high-energy physics}

High-energy physics involves a complex web of mathematical and technical knowledge, whether it concerns the details of the often abstract underlying theories, the behavior of the instruments that are assembled within sophisticated experiments, statistical notions for the analysis of the data derived from these experiments, etc. As a result of this complexity, there is a strong division of labor within high-energy physics, and we can even distinguish two different groups within the theorists themselves.  While ``pure'' theorists (we will call them ``theorists'', in accordance with the terminology within the field) are driven by ``the abstract elaboration of respectable theories'', phenomenologists (the second kind of theorists) are often more concerned with ``the application of less dignified models to the analysis of data and as a
guide to further experiment'' \citep{pickering1984constructing}, or at least more concerned with experimental consequences rather than with high theory. This division is itself strong enough that these two kinds of physicists can generally receive different training and diverge early in their careers, although some physicists -- usually prominent ones -- have expertise in both these domains and are able to sustain exchanges between the two. Therefore, in the present paper, we will make the following claim:

\begin{quote}
    \textbf{Claim 1}: Over the historical range considered (1980--2020), categories ``theory'' and ``phenomenology'' in high-energy physics should be regarded as distinct subcultures with their own bodies of knowledge, ontologies and methodologies, and which are carried out by different people.
\end{quote}

 It is not controversial in itself that ``theory'' and ``phenomenology'' are different matters in \gls{hep}; these are now distinct categories within the \gls{hep} literature and it is not uncommon for physicists to label themselves as ``theorists'' or ``phenomenologists'' depending on their specialization. However, our claim goes further by stating that the nature of their work is \textit{so} distinct that it should not be assumed \textit{a priori} that they can sustain fruitful connections; per Galison, we should not expect \textit{a priori} that subcultures are bound to cooperate flawlessly under any circumstance; we should instead remain open to the possibility that they may fail to produce constructs of shared value within the contexts of their respective enterprises. There may not even be one single overarching goal that is equally shared and sought after by \gls{hep} theorists and phenomenologists, and it is even less certain that their respective methods should equally contribute to achieving their goals at any time\footnote{\citealt{Galison1995} provides a distinction between two kinds of theorists similar to the one we propose to make here, resting on the recognition that these two groups rely on very different sets of constraints as guides towards theoretical progress.}. In the following subsection, we will propose that supersymmetry exemplifies the contingent ability of high-energy physicists to coordinate their respective methods and goals in a successful way. It does so because the story of supersymmetry is that of a partial failure, rather than that of a total success. Although successful cases of cross-fertilization across fields are valuable to illustrate the notion of trading zones, that science (and even physics itself, as Galison claims, against a symbiotic view of theory and experiment) is disunified is better exemplified by those cases where scientific cultures attempt and fail to establish coordination. The dramatic story of supersymmetry provides such an example.


\subsubsection{Supersymmetry as a trade-able good between theory and phenomenology}

Supersymmetry is a symmetry that relates the two fundamental kinds of particles that arise in nature: fermions and bosons. It was postulated simultaneously and independently by several physicists in the early 1970s, who were each motivated by very different goals\footnote{
For a history of early supersymmetry, see \citealt{kane2000the}.}. Supersymmetry rapidly gathered substantial attention from the theoretical community. The reasons were manifold, but they were clearly theoretical rather than empirical, as early reviews of the topic show\footnote{\citealt{Fayet1976cr,Freedman1978zi,Taylor1983su} provide a good overview of the main arguments for supersymmetry in its early days, all of which are highly theoretical.}. First, symmetry principles play a fundamental role in High-Energy physics, and supersymmetry was an especially attractive symmetry because of its peculiar properties
. Second, supersymmetry can naturally give rise to gravity, as was observed by \citet{Volkov1973}, suggesting that it could lead to a consistent theory of quantum gravity. This feature of supersymmetry gave birth to an entire research program, ``supergravity'', which then spanned several decades\footnote{Later on, supersymmetry proved even more interesting to theorists, by improving the consistency of string theory, and by supporting the conjectured AdS/CFT correspondence, yet another major development in quantum gravity research.}. Third, while quantum field theory is prone to mathematical difficulties due to divergences appearing in the perturbative calculations of certain quantities, in many instances such infinities were suppressed in supersymmetric theories.

However, as appealing as it was to theorists, supersymmetry posed a number of empirical difficulties. First, supersymmetry establishes a symmetry between bosons and fermions; and yet, at first it was not  at all clear which of the bosons and fermions should have been related to each other by this symmetry. Moreover, if supersymmetry were perfectly realized in nature, the particles it relates should have identical masses, which was also in contradiction with the data. This contradictory situation was well summarised by \citet{Witten1982de} in his \textit{Introduction to supersymmetry}: 

\begin{quote}
    [Supersymmetry] is a fascinating mathematical structure, and a reasonable extension of current ideas, but plagued with phenomenological difficulties. [...] Supersymmetry is a very beautiful idea, but I think it is fair to say that no one knows what mysteries of nature (if any) it should explain.
\end{quote}

Still, efforts to incorporate supersymmetry into a theory consistent with the data were undertaken over several years, and they culminated in what is now called the \gls{mssm} \citep{Fayet1976cr,Dimopoulos1981}. The \gls{mssm} is the result of reconciling the achievements of the \gls{sm} (the best theoretical account available at the time and still today) with the requirement of supersymmetry. This, however, has very undesirable consequences. Compared to the \gls{sm}, the \gls{mssm} introduces 105 additional unspecified parameters, so that supersymmetry can accommodate a large range of observations and has little predictive power in general \citep[p.~1]{Parker1999}. In particular, although supersymmetry predicts the existence of many new particles (the ``superpartners''), there is \textit{a priori} little chance  that these particles will have just the right properties to be discoverable in experiments. If not, supersymmetry may be of high value to theorists (because of its mathematical properties, and its promise to achieve a coherent account of quantum gravity), while being of low value to phenomenologists who are interested in building predictive models that can lead to the discovery of new particles or phenomena\footnote{Supersymmetry suffers from other disadvantages. For instance, many parameters of the theory imply certain phenomena to extents that have not been observed, e.g., baryon and lepton number violation, or flavor changing neutral currents \citep[201--209,235--240]{weinberg1995the}, which requires \textit{ad-hoc} explanations as to why, although allowed by the model, these mechanisms do not occur in nature.}.

Yet, in 2011, supersymmetry was perceived across the field as the theory beyond the \gls{sm} that was most likely to manifest itself in experiments \citep{Mttig2019,Mttig2020a}. Arguably, the reason why it became highly credible and valuable to phenomenologists as well, was that it could solve the so-called ``naturalness'' problem of the standard model on the condition that it was discoverable. 
In parallel to these developments around supersymmetry, there was indeed increasing recognition that an explanation was required as to why the the mass of the Higgs boson (an important piece of the Standard Model) could be many orders of magnitude below the mass scale at which the unification of forces is assumed to take place. 
It was also realized that supersymmetry could provide an answer to this ``naturalness''  problem (\citealt{Weinberg1978,Veltman1980,Witten1982de}), \textit{but} only as long as the masses of the superpartners (the particles predicted by supersymmetry) are not too high, so that they should be discoverable in future experiments\footnote{One can put other constraints on the \gls{mssm}, by requiring that supersymmetry explains dark matter, or that it ensures the convergence of the ``couplings'' that measure the strength of the fundamental forces at different length scales, which suggests it should play a role in the unification of these forces. However, as \citealt{Giudice2004} put it, ``the unification and dark-matter arguments [for supersymmetry] are not in general sufficient to insure that new physics be within the LHC discovery reach, contrary to the naturalness criterion''.}. In light of this, supersymmetry became of very high value to phenomenologists and experimentalists as well, rather than just a mathematical toy for the theorists to play with\footnote{The naturalness argument also provides a ``narrative'' that connects what theorists are concerned with (the details of the theories at energy scales unattainable in the experiment) to what experimentalists can probe. As \citet[p.~76]{Borrelli2015} puts it, ``the strength of the naturalness narrative is largely due to its flexibility, which allows it to become a unifying factor in the high-energy community and to bridge the gap between theorists and experimenters''  }.

\begin{figure*}
    \centering
    \begin{tikzpicture}[scale=0.75]
     \node (a) at (0, 0) {};
     \node (b) at (0, 6) {};
     \node (c) at (4, 0) {};
     \node (d) at (4, 6) {};

     \node (e) at (-3.25, 3) {};
     \node (f) at (7.25, 3) {};

     \node[above of=b,font=\bfseries,align=right,xshift=-3em] (th) {Theory}; 
     \node[above of=d,font=\bfseries,align=left,xshift=+3em] (ph) {Phenomenology}; 
     
     \node[below left of=th,align=right,xshift=-1em,yshift=-1em] (th_goals) {Unification?\\ Quantum Gravity?}; 
     \node[below right of=ph,align=left,xshift=+1em,yshift=-1em] (ph_goals) {``New physics''?};

     \draw[color=blue] (-0.25,1.5) rectangle (4.25,4.5) node[align=center,pos=0.5] {Supersymmetry};
    
     \draw[->,very thick]  (a.north east) to [out=45,in=-45] (b.south east);
     \draw[->,very thick]  (c.north west) to [out=135,in=-135] (d.south west);

     \draw[->,color=red,thick]  (e) -- (0.25,3) node[midway,above] {Naturalness};
     \draw[->,color=red,thick]  (f) -- (3.75,3) node[midway,above] {Naturalness};
     
     \node (g) at (0.125,0.875) {};
     
         \node[left of=g,align=right,xshift=-2em,fill opacity=0, text opacity=1] {Renormalizability,\\Symmetries,\\Consistency...};
         
         \draw[->,color=black] (0.125,0.875) -- (0.375,0.625);
         \draw[->,color=black] (0.875,0.125) -- (0.625,0.375);
        
         \node (h) at (3.875,0.875) {};
         
         \node[right of=h,align=left,xshift=3.25em,fill opacity=0, text opacity=1] {Collider/astrophysical data,\\generic models\dots};
         
         \draw[->,color=black] (3.875,0.875) -- (3.625,0.625);
         \draw[->,color=black] (3.125,0.125) -- (3.375,0.375);

     \end{tikzpicture}
    \caption{\textbf{Supersymmetry in the trading zone between theory and phenomenology.} Theorists and phenomenologists have different aims and methodologies, and whether they can both positively appraise a particular construct is not guaranteed. In the case of supersymmetry, it is the naturalness requirement that ensures that the \gls{mssm} is so valuable to both subcultures. As a result supersymmetry enhances a trading zone between these two cultures.  }
    \label{fig:mssm}
\end{figure*}
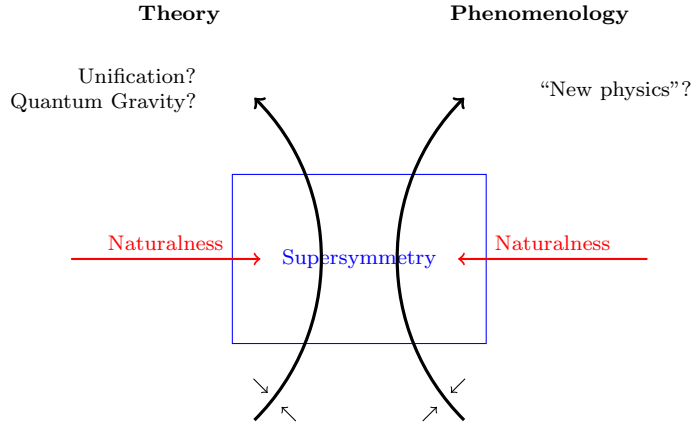

This situation is summarised in Figure \ref{fig:mssm}. As theorists work out a path towards their goals (e.g., the unification of forces, or the formulation of a consistent theory of quantum gravity), they rely on theoretical heuristics such as renormalizability, symmetry principles, consistency requirements, etc. \citep{Galison1995}. In that context, supersymmetry emerges as a very valuable concept. Phenomenologists, on the other hand, try to work out a path towards the discovery of ``new physics'' (evidence for new phenomena unaccounted for by the \gls{sm}) by relying instead on more generic models and constraints derived from experimental data (e.g. from particle colliders or astrophysical observations). It is the naturalness requirement that makes supersymmetry valuable to phenomenologists as well, by strengthening the belief that supersymmetric particles should have masses that are low enough to be discoverable. In this way, supersymmetry effectively enhances the ``trading zone'' between theorists and phenomenologists: both communities can acknowledge its value in spite of the vast differences in their aims, methods, and objects of inquiry.

It is now time to introduce the last (but not the least) player in our drama: the \gls{lhc}. Operating since 2010, the \gls{lhc} is the largest physics experiment ever built. By performing particle collisions at the highest energies ever achieved, it promised to discover supersymmetric particles, provided that they had the properties prescribed by the naturalness problem that supersymmetry should solve. However, no such discovery has been made, which suggests that the ``naturalness problem'' was unwarranted \citep{Giudice2017}. If there is no naturalness problem, then, supersymmetry is left unconstrained again; there is no guarantee that supersymmetric particles will ever be discovered; and its phenomenological value plunges back to the depths from which it surfaced. Therefore we will put forward the following claim, which will also be evaluated in the present paper:

\begin{quote}
    \textbf{Claim 2}: Supersymmetry occurs in a variety of partially independent contexts within high-energy physics, some of which belong to ``theory'' and some of which belong to ``phenomenology'', and these applications of supersymmetry have responded differently to the \gls{lhc}'s failure to find supersymmetric particles.
\end{quote}

Furthermore, we hypothesized that supersymmetry should be losing its ability to sustain trades between theory and phenomenology. Therefore, we will evaluate the following claim:
 
\begin{quote}
    \textbf{Claim 3}: Supersymmetry sustained trades between theory and phenomenology in high-energy physics, until it was challenged by the \gls{lhc}'s failure to observe the particles predicted by supersymmetry.
\end{quote}

If theorists and phenomenologists fail to share a similar appraisal of supersymmetry, then this may pose a serious problem for the field: this would imply that theorists' research programs can persist despite their low value to phenomenologists, and conversely that experimental input has little to offer to theorists; if that is the case, then the unity of high-energy physics would indeed be fragilized. Therefore, addressing  claims 1-3 (1--that theory and phenomenology are partially autonomous subcultures of high-energy physics; 2--that supersymmetry arises in distinct, autonomous contexts, which responded differently to the absence of supersymmetric particles at the \gls{lhc}; and --3-- that the value of supersymmetry for bridging together subcultures of physics has decreased as a result of the failure of phenomenological supersymmetry) should contribute to answering the questions of what makes and unmakes unity in \gls{hep}.

\subsection{Towards a quantitative assessment of subcultures and trades}
\label{section:quantitative}

In the following, we propose an array of quantitative methods implementing several dimensions of Galison's framework for addressing the plurality of science, which evaluate the claims put forward above. To this end, we will rely on authorship data (for investigating the social entrenchment of theory and phenomenology as distinct subcultures), semantic analyses (for investigating the linguistic divide between these subcultures as well as the plurality of supersymmetry research), and citation data (in order to locate ``trading zones'' within the field). To our knowledge, this is the first attempt to implement Galison's framework into a quantitative analysis of scientific literature. Of course, the plurality of science and the coordination between scientific fields have already been addressed quantitatively in numerous publications. In the context of physics research, for instance, \citet{Battiston2019} have evaluated the ability of physicists to publish in various subfields. In particular, they demonstrate that \gls{hep} physicists are among the most specialized physicists (i.e., they have a high probability of publishing only in their primary subfield), although their work does not distinguish between the various kinds of high-energy physicists, which will be done in the present paper. There remains to address the linguistic component of the divide between these subcultures, in particular theory and phenomenology, and to this end we will propose a novel strategy based on semantic data (titles and abstracts of the literature).

As for the analysis of the plurality of supersymmetry-related research in \gls{hep}, we will develop a topic model approach in order to identify clusters of concepts that are most likely to be associated with supersymmetry in the literature, and we will explore the dynamics of supersymmetry research throughout time.

Finally, we will assess the intensity of trades between theoretical subcultures and locate the concepts that facilitate these trades. \citet{Yan2013} proposed a quantitative assessment of dependency relations between scientific disciplines based around a metaphor with international trade, by measuring quantities such as ``exports'', ``imports'', or ``self-dependence'' of various fields throughout time based on citation data. However, this work does not investigate what exactly allows these trades to happen, e.g., which concepts sustain them. This requires combining citation data with semantic information about papers' concepts, as achieved by \citet{Raimbault2019} who proposed measures of interdisciplinarity built upon such data. Similarly to \citealt{Yan2013}, we will assess the self-dependence of experiment, phenomenology, and theory in \gls{hep} based on the citation network. However, we will also evaluate the ability of different concepts (such as supersymmetry) to sustain trades across subcultures throughout time, by combining semantic and citation data.




More broadly, this work will add to quantitative studies of science literature, by helping to fill a gap that has come to the attention of the community.  As stressed by \citet{Leydesdorff2020,Kang2020,Bowker2020}, quantitative and qualitative studies of science have mostly diverged in their goals and ``world views'', urging the need to ``bridge the gap'' between them. We propose, therefore, a bridge connecting these two forms of scientific study. First, we demonstrate that quantitative methods can address questions raised by the philosophy, history, and sociology of physics. Moreover, we show that concepts from qualitative 
 science studies can give structure to quantitative methods, in line with the call by \citealt{Heinze2020} to inform quantitative analyses with ``middle-range theories'' (of which Galison's trading zones are an example). As a result, our methods are in principle meaningful in any context where such a theory is valid -- whenever scientific cultures attempt to achieve coordination --, much beyond the case-study proposed in this paper.

\section{Methods}\label{section:method}

\subsection{Data\label{section:data}}


Our data consists of the scientific literature on high-energy physics and the semantic, authorship, and citation information that it entails which is of interest for our questions.
 
The data were retrieved using the Inspire HEP database \citep{InspireAPI}. Inspire HEP is a platform dedicated to the HEP community and is maintained by members of CERN, DESY, Fermilab and SLAC. It aggregates publications from the \gls{hep} literature, and maintains a list of institutions and collaborations involved in the community, while also publishing job offers. It replaced Spires in 2012\footnote{``Physicists, start your searches: INSPIRE database now online'', \textit{Symmetry}, May 24th 2012, \url{https://www.symmetrymagazine.org/breaking/2012/05/24/physicists-start-your-searches-inspire-database-now-online}
}.

The database is fed by an automatic aggregator that retrieves articles from multiple sources\footnote{Melissa Clegg, ``INSPIRE Content Sources
'', May 30th 2020 (\url{https://help.inspirehep.net/knowledge-base/inspire-content-sources/})} including a number of databases (including the Astrophysics Data System, arXiv, etc.), research institutions (CERN, DESY, Fermilab, IHEP, IN2P3, SLAC), and scientific editors such as the American Physics Society or Springer.

Inspire then aggregates data from these sources with automated crawlers, and it performs manual curation for completion or error-correction\footnote{Stella Christodoulaki, ``Content Policy'', March 4th 2020, \url{https://help.inspirehep.net/knowledge-base/content-policy/}}, including author name disambiguation. This database has a strong yet untapped potential for quantitative analyses. However, only contents related to HEP are subject to a systematic effort of collection and curation, and the data should be used preferably in analyses which scope is limited to HEP, thereby making it unsuitable for, e.g., studying interactions between HEP and other fields of physics (e.g. condensed matter physics).

The database includes data about the contents of the literature (title, summary, sometimes keywords), the authors (name, unique identifier, institutional affiliations), dates corresponding to different events related to each paper, associated experiments, references of the articles. The only data pertaining to the contents of the articles that are consistently available are titles and abstracts. Articles are categorized according to a classification scheme compatible with that of the arXiv pre-print platform. This scheme includes categories such as \texttt{Theory-HEP}, \texttt{Experiment-HEP}, \texttt{Phenomenology-HEP}, \texttt{Astrophysics}, etc. Categories of papers published on arXiv.org are extracted directly from the platform (where they are defined by the authors, while being subject  to moderation and controls). Categories of papers not published on arXiv.org are now assigned manually by curators. Categories of papers inherited from the ancestor of Inspire HEP (Spires) and absent from arXiv.org were derived according to a mapping between Spires' classification and the current arXiv-based classification. In this paper, we rely mostly on three categories which entail most of the high-energy physics literature: \texttt{Theory-HEP}, \texttt{Phenomenology-HEP}, but also \texttt{Experiment-HEP}, which typically entails papers that report empirical results such as statistical analyses of experimental data. A portion of the articles between the years 1990 to 1995 were not categorized, which led to some issues with the data collection process, as described in Appendix \ref{appendix:collection}. For this reason, our longitudinal analyses will focus on later years, which does not prevent us from addressing our research questions. The analysis of subcultures spans over years 1980 to 2020. The years prior to 1980 could also have been interesting for this analysis as well, but the corresponding data was of lower quality.

\subsection{Social and semantic analysis of subcultures of high-energy physics}\label{section:method_subcultures}

The first claim that we seek to establish is that ``theory'' and ``phenomenology'' should both be regarded as distinct subcultures within physics. There are two components to subcultures: a linguistic one (they should have vocabularies that are distinct enough to signal complementary bodies of knowledge) and a social one (they should correspond to distinct groups of people). Therefore we will proceed twofold. First, we will demonstrate that theory and phenomenology manipulate vocabularies that are so distinct that we can predict with reasonable accuracy whether a paper belongs to one of these categories based on the words present in its abstract; our predictive model will then be used to unveil the ontological differences between these subcultures. Second, we will show that these categories from the literature are associated with different communities.

\subsubsection{The semantic divide between Theory and Phenomenology}

If it is possible to tell whether a paper is theoretical or phenomenological based on the words it contains, then this implies that these categories use partially distinct vocabularies -- i.e., that each of these two categories has its own ``language'' -- in a way that allows papers from one category to be distinguished from those from another. If that is the case, we can then examine the nature of the linguistic divide between theory and phenomenology in order to better understand their differences. In what follows, we apply this strategy using statistical methods, based on the classification of \gls{hep} literature provided by Inspire HEP\footnote{Since this classification relies on a manual assignment of the different categories, any potential linguistic divide between them cannot be the byproduct of some algorithmic bias}. Although we are more interested in the divide between ``theory'' and ``phenomenology'', we also include ``experiment'' (which Galison himself labeled as a subculture of its own) in our analysis in order to emphasize its differences with phenomenology.

In order to establish whether we can predict which articles $d$ belong to any of the categories $c \in \{$Experiment, Phenomenology, Theory$\}$, we will build a simple linear logistic regression using a bag-of-words as the predictive features. In this approach, the corpus is represented by a matrix $B=(b_{d,i}) \in \mathbb{R}^{D\times V}$ where $D$ is the amount of documents, $V$ is the size of the vocabulary, and $b_{d,i}$ is the number of occurrences of the word (or expression\footnote{We also include some n-grams in the model, i.e. expressions of several words, provided they follow certain pre-defind syntactic patterns (e.g. ``adjective +noun'')}) $i$ in the document $d$. This representation excludes a lot of semantic information that results from the knowledge of the ordering of the words and the structure of sentences within the documents; it is in line with our goal to find out whether the vocabularies of each category are so distinct that the mere presence or absence of certain words can be used to infer the category of a document. We perform a normalization of the bag-of-words prior to the regression, by applying the tf-idf transformation\footnote{For a definition of the tf-idf transformation, and information theoretic justifications of its relevance, see \citealt{Beel2015_tfidf_uses,Robertson2004_tfidf_justification}. We use scikit-learn's implementation of the inverse-document frequency transformation which is $\text{idf} = 1+\log(1/f)$ where $f$ is the fraction of documents in which a word occurs. It differs from the ``textbook'' definition $\log(1/f)$ because of the regularization term ($+1$).} to $(b_{d,i})$, resulting in a normalized bag-of-words which we will name $(b_{d,i}')$. More specifically, our predictive model is defined as: 

\begin{equation}
    \label{eq:logistic_bow_classifier}
    P(d\in c) = \mathrm{logit}^{-1}\left(\beta_c + \sum_{i=1}^V \beta_{ci} b'_{di}  \right)
\end{equation}

This model is then trained on $N=100,000$ articles of our database from 1980 to 2020 that belong to any of the following categories: \texttt{Experiment-HEP}, \texttt{Phenomenology-HEP}, and \texttt{Theory-HEP}, \footnote{The fit is performed with the scikit-learn python library \citep{scikit-learn} using L2 regularization.}. The vocabulary used in the regression is the $V$ expressions (n-grams, up to four word long) among those that belong to predefined syntactic patterns\footnote{We choose a subset of the syntactic patterns used to analyze the ACL anthology corpus in \citealt{omodei_tel-01097702}.}, that have the highest ``unithood'' as measured in \citealt{omodei_tel-01097702}\footnote{The ``unithood'' measures ``the degree of strength or stability of syntagmatic combinations or collocations'' \citep{Kageura1996}}. The size of the vocabulary $V$ is chosen to be a round number that is just high enough to reach about the maximum accuracy of the model, as evaluated on the test set (which consists in 10,000 articles not present in the training set). Then, the accuracy of the predictions of the model is evaluated using the same test set. The coefficients $\beta_{ci}$ are then analyzed in order to extract the words that are the most discriminatory between ``theory'' and ``phenomenology'', thus revealing the most salient differences. For that, we retrieve those expressions $i$ that maximize $\beta_{\text{th},i}-\beta_{\text{ph},i}$ and $\beta_{\text{ph},i}-\beta_{\text{th},i}$. Because of the inverse document frequency transformation applied prior to the regression, expressions that are more common are favored by this selection process.

\subsubsection{The social divide between Theory and Phenomenology}

What does it mean to say that theory and phenomenology have a ``demographic component'', as  \citet[p.~138]{galison1987how} puts it, regarding theory and experiment in \gls{hep}? It means that these categories of the literature are supplied by distinct groups of people, ``theorists'' and ``phenomenologists''. Therefore, we will investigate whether it is the case that experimental, phenomenological and theoretical papers are published by three distinct groups of physicists, such that these physicists usually contribute mostly to just one of these categories Again, ``experiment'', which is a paradigmatic example of subculture in Galison's view, is also included in our analysis. It will be useful to assess whether the distinction between phenomenology and theory is comparable to the distinction between theory and experiment (the one initially stressed by Galison).

Let $N_{ij}$ be the amount of articles co-authored by a physicist $i$ that belong to the category $j \in \{$theory, phenomenology, experiment$\}$, and $N_i$ the total amount of articles co-authored by $i$. Let us assume $N_{ij} \sim \mathrm{Binomial}(N_i, p_{ij})$, where $p_{ij}$ is the latent probability that a paper from physicist $i$ belongs to the category $j$\footnote{Since these categories are not mutually exclusive in our database (an article may belong to more than one of them), a multinomial process would not be a good fit.}. Since the researchers co-authored widely varying amounts of publications (ranging from a few papers to hundreds), we assumed that the latent probabilities $p_{ij}$ were described by the following model:

\begin{align*}
    N_{ij} &\sim \mathrm{Binomial}(N_i, p_{ij})\\
    p_{ij} &\sim \mathrm{Beta}(\alpha_j, \beta_j) \\
    \alpha_j,\beta_j &\sim \mathrm{Exponential}(1)
\end{align*}

The binomial process assumes that each physicist can be imputed a constant latent fraction of papers in each category. The beta prior is a flexible distribution over probabilities, which can be either unimodal or bimodal. The exponential prior over $\alpha$ and $\beta$ is agnostic regarding these two possibilities, and its exact shape does not significantly matter considering the amount of available data. Most crucially for us, this model allows to combine information from researchers with many papers and researchers with very few papers; for those with few papers, the estimation of the latent probabilities is more influenced by the shape of the beta distribution. The model was fitted to 2500 researchers randomly sampled among those with more than 3 publications in HEP for 1980-2020. In order to evaluate the social entrenchment of these categories, we verify that most physicists contribute mostly to just one of these categories.

\subsection{Assessing the plurality of supersymmetry research with topic models}\label{section:method_plurality}

Our second claim pertains to the plurality of supersymmetry research. In this section, we present our methodology for assessing the plurality of supersymmetry related research, by recovering the contexts, i.e., the topics in which supersymmetry occurs, and by evaluating the extent of their independence, and how they responded to the results of the \gls{lhc}. More broadly we provide a methodology for investigating scientific ``objects'' akin to ``boundary objects'' in that they are ``plastic enough to adapt to local needs and constraints of the several parties employing them'' \citep[p.~393]{Star1989}, by unveiling the plurality and autonomy of the contexts in which such objects may arise.

\subsubsection{Model}

In order to evaluate in which contexts supersymmetry arises within the high-energy physics literature, we have chosen to subdivide the literature into sub-topics using an unsupervised probabilistic topic model, namely the Correlated Topic Model (CTM, \citealt{Blei2007}). We do not use conventional classifications such as the \gls{pacs} codes from the \gls{aip}, because they were not available for the whole dataset -- \gls{pacs} codes were only available starting from 1995, and only for a subset of the papers, which may not be representative of the whole. Besides, \gls{pacs} codes are too numerous (more than 5000 categories)\footnote{``Full list of PACS numbers'', \textit{
Physics-Uspekhi}, \url{https://ufn.ru/en/pacs/all/}} for our purposes. Therefore, we opted to extract the topics in the literature using unsupervised topic models instead. 

Probabilistic topic models generally assume that each document of a corpus is a mixture of variable proportions of a certain amount of topics, each of these topics having their own vocabulary distribution. When trained on a corpus, such models simultaneously learn the ``topics'' in the corpus (and their vocabulary), as well as the relative contribution of each topic to each document of the corpus. These models have demonstrated their ability to capture the semantic information contained within the scientific and academic literature, as shown in previous work\footnote{Notable examples are \citealt{Nichols2014,Hall2008,Griffiths2004}; see \citealt{Malaterre2022} for a more recent application in the context of History and Philosophy of Science, and \citealt{Allen2022} for an assessment of the potential and limitations of these methods in the field.}, even from abstracts alone \citep{Syed2017}; as a result this technique has seemingly taken precedence over network-based semantic maps \citep[Figure~1]{Leydesdorff2016}.  Although co-occurrence networks may have more conceptual bearing in the STS tradition, we have preferred topic models for their intrinsic ability to capture the polysemy of certain words (e.g., ``supersymmetry''), in terms of the probabilities that such words can arise in different contexts (i.e. topics).

In particular, we have chosen the Correlated Topic Model (CTM) for its ability to capture correlations between topics. In this model, the contribution of a topic $z$ to a document $d$, $P(z|d)$, is assumed to be drawn from a hierarchical model involving a correlated multivariate distribution \citep{Blei2007}:

\begin{align}
    \label{eq:ctm_prior}
    \vec{\beta}_{d} &\sim \mathcal{N}(\vec{\mu}, \Sigma) \\
    P(z|d) &= \dfrac{\exp \beta_{d,z}}{\sum_{i=1}^k \exp \beta_{d,i}}
\end{align}

Through the covariance matrix $\Sigma$, the CTM model is able to learn correlations between topics, and therefore to account for the fact that some topics are more likely to occur together within one document. Moreover, our intuition is that using CTM allows the derivation of more realistic topic-distribution for short texts such as abstracts, for which the small amounts of words only moderately inform the prior topic distribution. Most importantly, this model allows us to directly assess the level of independence between the topics derived by the model, which is important for assessing the autonomy of the contexts in which supersymmetry arises.

The model is trained on $N=120,000$ articles randomly sampled from those between 1980-2020 that belong to any of the categories \texttt{Theory-HEP}, \texttt{Phenomenology-HEP}, \texttt{Experiment-HEP}, and also \texttt{Lattice} (a theoretical approach to HEP, with ties to both theory and phenomenology, and in which we expected supersymmetry to potentially arise as well). The procedures for extracting the input vocabulary and for choosing the hyper-parameters are described in detail in appendices \ref{appendix:data_selection} and \ref{appendix:hyper_parameter} respectively. Two methodological contributions can be highlighted. First, we included informative n-grams matching pre-defined syntactic patterns in the vocabulary in order to preserve more semantic information. Second, we made a prudent and balanced use of perplexity and topic coherence measures in order to recognize the advantages and limitations of both these kinds of measures for assessing the quality of topic models and choosing the best hyper-parameters. The procedure resulted in the extraction of 75 topics.

\subsubsection{Interpretation and validation}

Once the model was trained, we manually assigned a label to each topic, by inspecting and interpreting their top-words and the categories from the \gls{pacs} classification of the physics literature that were most correlated to each topic\footnote{We used pointwise mutual information (see equation \ref{eq:pmi_expression}, appendix \ref{appendix:validation}) as the measure of correlation.}. Informing our interpretation of each topic with these correlations rather than the sole top-words help overcome issues associated with the interpretation of fat-tailed topic-word distributions based on a handful of top-words \citep{Chang2009,Allen2022}.  We failed to provide a meaningful label for some topics, but this had little impact on the rest of the analysis. Finally, in order to assess the meaningfulness of the metrics produced by the model (the document-topic distributions and the topic-word distributions), we performed an additional validation procedure using the \gls{pacs} classification of the literature and the input of independent experts (see appendix \ref{appendix:validation}). 

In section \ref{section:application_plurality}, the model is applied to a number of tasks: the evaluation of the contexts (i.e. topics) in which supersymmetry occurs in the literature, the extent of the correlation between these contexts and finally the trends in research involving supersymmetry since the start of the \gls{lhc}.

\subsection{Locating trades across scientific cultures}\label{section:method_trading_zone}

In this section, we elaborate a longitudinal methodology for locating trades between scientific cultures, which we use to assess the ability of supersymmetry to enhance trades between the theoretical and phenomenological cultures of \gls{hep} throughout time. Trading zones can manifest themselves in a myriad of ways, some of which are readily prone to a quantitative analysis. 
For instance, citing the example of quantum chromodynamics, a theory of the strong interaction, Galison notes that ``the contact between the experimenters and the phenomenological theorists had grown to the point where Andersson [a theorist] and Hofmann [an experimentalist] could coauthor a \textit{Physics Letter}'' \citep[p.~655]{galison1997image}. In that sense, a paper co-authored by scientists from different cultures is indicative of a trading zone, such that co-authorship data can in principle be used to probe trades across scientific cultures. Another manifestation of trading zones can be found in the citation network, which encodes exchanges of knowledge across publications, and sometimes across subcultures. Indeed, that a phenomenological publication, for instance, cites a theoretical paper, indicates that phenomenologists can acknowledge the value and significance of certain theoretical constructs (that are present in this specific paper) in their enterprise.  Although in principle both the citation networks and the collaboration networks could be used for our purpose, the present analysis will rely on the former. Indeed, the citation graph preserves more information about the directionality of the exchanges involved, thus supporting the trade metaphor in \citealt{Yan2013}. Intuitively, it is also less vulnerable to non-epistemic factors as is the case with authorship (e.g. physicists authoring papers they did not contribute to as is frequent in large collaborations in the field). In addition, for validation purposes, we show  in Appendix \ref{appendix:phenomenology_centrality} (Fig. \ref{fig:cites_matrix}) that the citation network can indeed reveal the relative autonomy (self-reliance) of \gls{hep} subcultures, but also the special role of the phenomenological subculture in sustaining the unity of \gls{hep} by channeling trades across theory and experiment (which hardly communicate directly otherwise). This further supports the use of the citation graph use as a means of locating trades.

In order to assess the ability of supersymmetry to facilitate trades between theorists and phenomenologists, we develop a method that combines two important aspects of Galison's trading zones: their locality and their linguistic component (the ``inter-language''). In particular, we look for scientific concepts that are most likely to be involved in trades between these subcultures throughout time. To this effect we perform the analysis on a subset of the citation graph, such that the nodes are limited to theoretical and phenomenological papers, excluding cross-listed papers (those that belong to both these categories). For each of these two theoretical cultures, we derive a list of informative keywords from the abstracts of the papers by extracting $n$-grams ($n\geq 2$) matching certain syntactic patterns. We retain the top $N$ keywords (sorted by decreasing unithood) such that at least 95\% of the abstracts contain at least one of the $N$ keywords; this yields $N=1370$ keywords specific to the phenomenological culture and $N=1770$ keywords specific to the theoretical culture. From this we derive a bag of words $b_{ik}$ for each publication such that $b_{ik}=1$ if keyword $k$ is present in abstract $i$, and $b_{ik}=0$ otherwise. We then evaluate the probability that the keyword occurs in an abstract given the paper is involved in a trade between a theoretical and a phenomenological paper at a time $t$, which we write $P(b_k=1|\text{trade}_{i\to j},t)$. We consider trades in both directions: phenomenological papers citing theoretical papers (th$\to$ph), then in a second time theoretical papers citing phenomenological papers (ph$\to$th). To what extent supersymmetry helps sustain the trading zone between these theoretical cultures is roughly measured by $P(b_k=1|\text{trade},t)$ for those keywords $k$ related to supersymmetry. In this analysis, we explore citations appearing in papers published between $t=2001$ and $t=2019$ (covering similar range prior and after the start of the \gls{lhc}), and we include all cited papers published from 1980 onwards; it is unlikely that recent papers would cite publications from before 1980. However, since cross-listed papers, which we excluded, have become much more common in the database starting from 2010 for spurious reasons (a change in the classification procedure), we ran a separate analysis in order to assess the robustness of our results. In this second analysis, we included cross-listed papers and assigned them only one category based on their authors' primary subfield (the subfield to which they contribute the most). We found both analyses to produce similar results. In the following we report the results obtained by excluding cross-listed papers.

\section{Results}
\label{section:application}

\subsection{Theory and phenomenology as distinct subcultures}\label{section:application_subcultures}

Let us now examine our first claim that ``theory'' and ``phenomenology''  should be regarded as distinct subcultures within high-energy physics. The claim requires that these categories mobilize distinct bodies of knowledge which manifest themselves through distinct vocabularies. As shown in Table \ref{table:categories_bow_prediction}, it is indeed possible to predict with reasonable accuracy whether a paper belongs to either one of these categories based on the vocabulary in its abstract. The accuracy is higher than 90\% for ``theory'' and reaches 86\% for ``phenomenology'', far above what one would obtain from assigning the most probable class irrespective of the contents, purely based on their frequency. This conclusion holds throughout the whole historical period considered (see Appendix \ref{appendix:stability}). This supports the existence of a linguistic divide between these two theoretical cultures over the years 1980 to 2020.

\begin{table}[h]
\centering
\begin{tabular}{@{}lllll@{}}
\toprule
               & Theory & Phenomenology & Experiment &  \\ \midrule
Model accuracy & 91\%   & 86\%          & 92\%       &  \\
Baseline       & 55\%   & 51\%          & 84\%       &  \\ \bottomrule
\end{tabular}
\caption{\textbf{Accuracy of the model for predicting which categories \gls{hep} papers belong to}. The precision of the model for each category is estimated based on the test corpus. For reference, the accuracy of a naive model that assigns the most likely class irrespective of any information about the papers is given (baseline). The size of the vocabulary used for the predictions is set to 500 words and expressions. }
    \label{table:categories_bow_prediction}
\end{table}

Our model also unveils the expressions that are most capable of discriminating between theory and phenomenology, as shown in Table \ref{table:specific_pheno_vocabulary_th_ph}. One striking difference between theory and phenomenology appears to be the importance of space-time related concepts in theory (``space-time'', ``geometry'',  ``manifold'', ``dimension'', ``coordinate'', etc.). The objects (entities) of interest also differ, which signals an ontological divergence: on the pure theory side, ``black hole[s]'' and ``strings'' are prominent entities, while particles (``quark'', ``neutrino'', ``gluon'', ``hadron'', ``nucleon'', etc.) belong to the realm of phenomenology. Among those terms most specific to phenomenology but absent in pure theory, we also  find the notions of model (``mssm'', ``standard model''), and effective field theories (``effective theory'', ``chiral perturbation theory'') which are approximate theories emerging from more fundamental theories. Moreover, the mention of ``experimental data'' is a distinctive feature of phenomenology: theory is not directly committed to establishing a connection with empirical results. Interestingly, one aspect of supersymmetry (the \gls{mssm}) appears as markedly phenomenological, while ``supergravity'' is specifically theoretical.

\setlength\extrarowheight{2pt}\input{Table2.tex}

Similarly, keywords that discriminate the most between experiment and phenomenology are shown in Table \ref{table:specific_pheno_vocabulary_exp_ph}. They confirm the theoretical ( ``model'', ``scenario'', ``effective theory'', ``implication'') and computational (``estimate'', ``approximation'',  ``contribution'', ``numerical result'', ``correction'') nature of phenomenology, as opposed to the empirical, ``fact-based'' dimension of experiment (``measurement'', ``search'', ``experiment'', ``event'', ``result'', ``evidence'', ``data'').

\setlength\extrarowheight{2pt}\input{Table3}

\begin{figure*}
    \centering\includegraphics{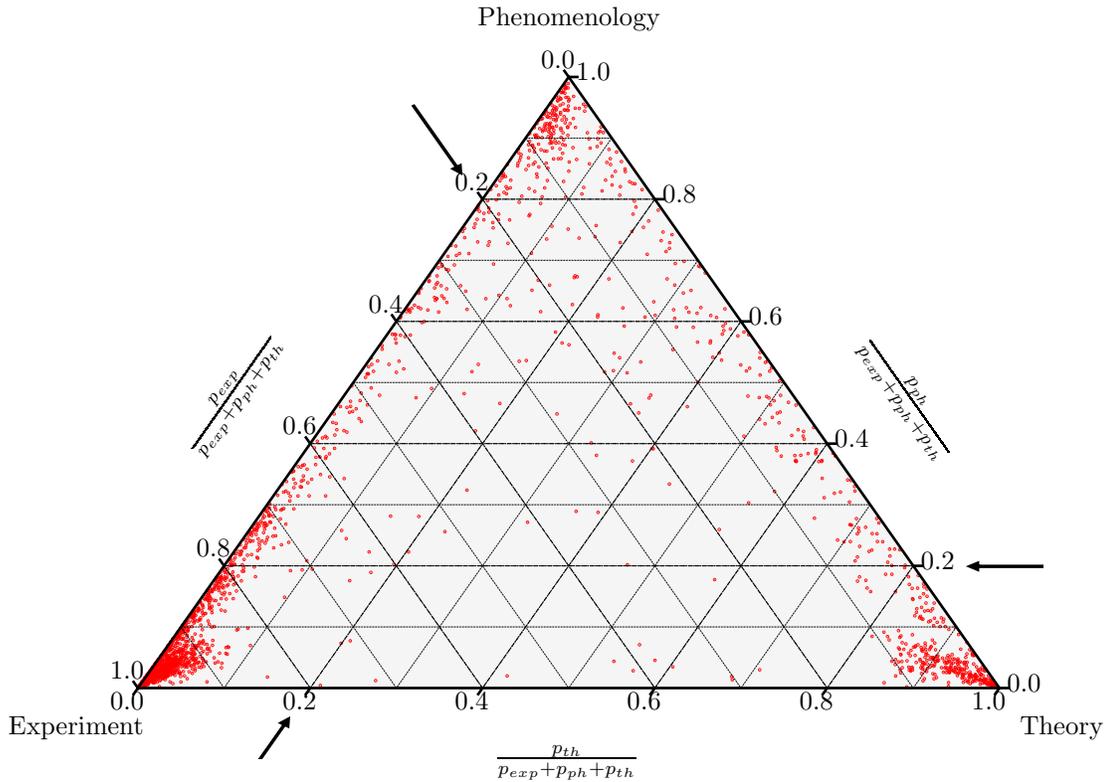}
    \caption{\textbf{Relative fraction of articles from any of the categories ``Experiment'', ``Phenomenology'' and ``Theory'', for 2,500 \gls{hep} physicists.} Each physicist among those sampled is represented by a red dot on the diagram, positioned according to the estimate of $(p_{i,exp},p_{i,ph},p_{i,th})$, the probability that any of his articles belong to those three categories. The dashed lines, along the direction of the arrows, form a grid along which one can read the relative importance of each category for every physicist  ($p_{ij}\over \sum_{k}p_{ik}$). Physicists near the vertices of the triangle contribute almost exclusively to one category; those near an edge contribute quasi-exclusively to two categories. Most physicists are located near a vertex, thus contributing to mostly one category.}
    \label{fig:ternary_categories}
\end{figure*}

What about the ``demographic component'' of the divide between theory and phenomenology? Do these categories have social counterparts? The results of our social analysis are shown in Figure \ref{fig:ternary_categories}. 
Figure \ref{fig:ternary_categories} is a ternary diagram in which each red dot represents a physicist and is positioned according to the relative prevalence of each category (among experiment, phenomenology and theory) among the papers they authored or co-authored. The majority of the dots are clustered near vertices, which means that most physicists dedicate themselves to mostly one of these categories. In particular, the inner part of the ternary diagram, which corresponds to physicists with balanced contributions to each category, is almost empty. We do find that some authors are scattered along the experiment-phenomenology edge and the phenomenology-theory edge; still, our results suggest that the category of phenomenology does feature a ``demographic'' counterpart as well, although it is more porous than experiment or pure theory. Therefore, phenomenologists do, to some extent, constitute a social group distinct from that of theorists (and experimentalists); however, phenomenology seems to play a special role in sustaining some form of cooperation between experimentalists and theorists. Overall, we find that 81\% of high-energy physicists publish more than 80\% of their papers in just one of these categories, which is clear evidence of specialization.

Our quantitative analysis supports our claim that, at least over the years 1980 to 2020, theory and phenomenology should be regarded as distinct subcultures with partially distinct languages. Consequently, strategies ought to be devised for them to properly communicate and coordinate their efforts, as long as physicists believe it to be necessary or worthwhile. It follows that their unity should not be assumed; instead, why a trading zone may be successfully worked out remains to be explained. Before we turn to the ability of supersymmetry to sustain the coordination between these subcultures, we will address the plurality of supersymmetry research itself.

\subsection{The plurality of supersymmetry}
\label{section:application_plurality}

In this section we apply our methods to address our second claim regarding the plurality of supersymmetry research and the recent decline in phenomenological supersymmetry research as a response to  \gls{lhc} results.

Topic models are able to link one word to several topics, thus allowing us to unveil different aspects of supersymmetry, i.e. different contexts\footnote{Like \citet{Allen2022}, we caution that these ``topics'' may not be as coherent as the common understanding of the word may suggest and that they should really be understood as different ``contexts'', although we use both terms interchangeably below.} in which this concept may occur. For three words $w$ that explicitly refer to supersymmetry (``supersymmetry'', ``supersymmetric'', ``susy''\footnote{Short for ``supersymmetry''}), we evaluated the probability $P(z|w)$ that these words occur in the context of a topic $z$ according to:

\begin{equation}
    P(z|w) = \dfrac{P(w|z)P(z)}{P(w)}
\end{equation}

\begin{figure*}
    \centering
    \includegraphics{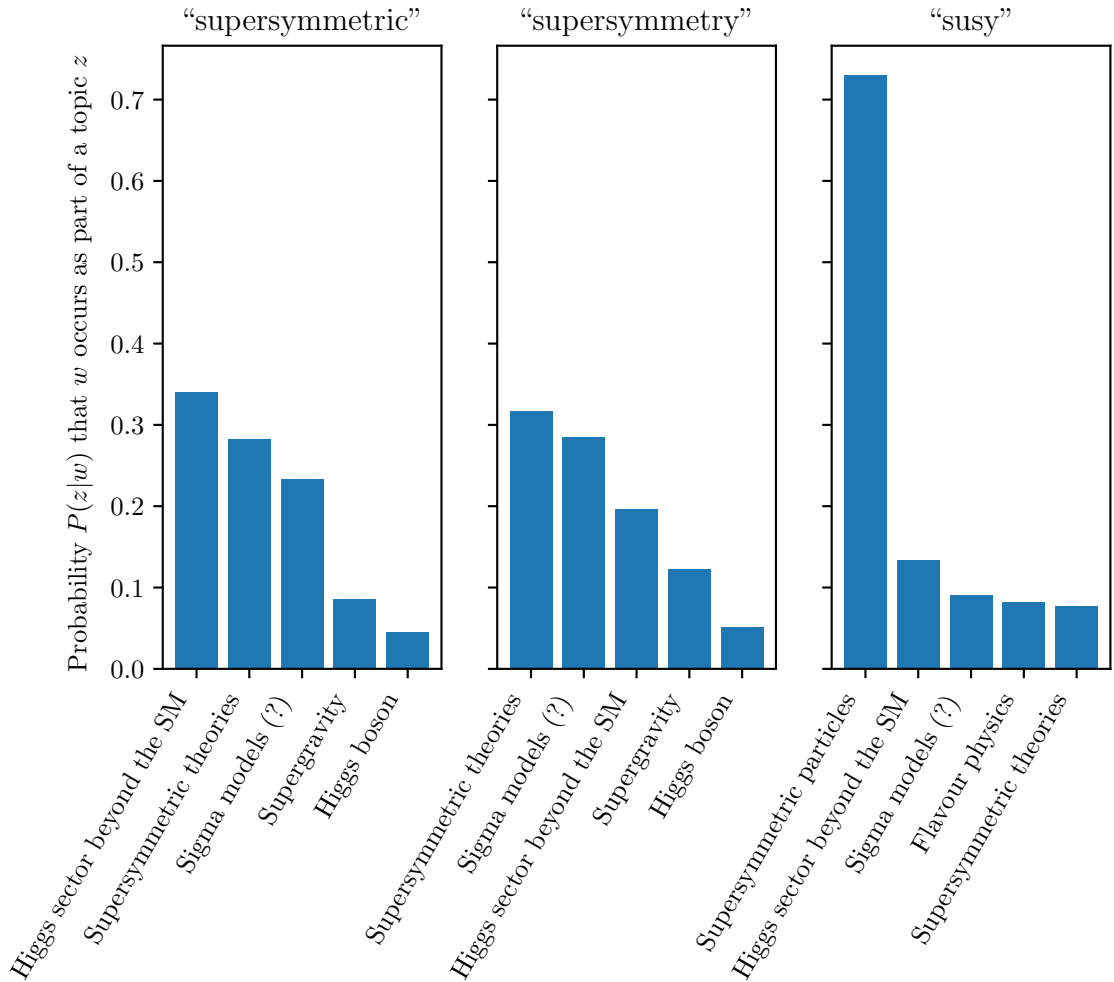}
    \caption{\textbf{The many uses of supersymmetry}. For three terms $w$ refering to supersymmetry (``supersymmetric'', ``supersymmetry'', and ``susy''), the five topics $z$ that are most likely to have led to their occurrence and their respective conditional probability $P(z|w)$ are shown. ``Supersymetry'' and ``supersymmetric'' have similar distributions, and mostly occur within theoretical topics. ``Susy'''s topic distribution is much more peaked, and most often occurs within phenomenological topics.}
    \label{fig:susy_usages}
\end{figure*}

Where $P(w|z)$ is frequency of the term $w$ within the topic $z$, $P(z)$ is the marginal probability of topic $z$, and $P(w)$ is the overall term-frequency of $w$. The five most probable topics for each of the words ``supersymmetry'', ``supersymmetric'', and ``susy''  are shown in Figure \ref{fig:susy_usages} (for the other topics, the probability $P(z|w)$ for $w \in \{$ ``supersymmetry'', ``supersymmetric'', ``susy'' $\}$ is residual). We can see that each of these terms may indeed occur in relation to a variety of topics: ``supersymmetric theories'' (which entail supersymmetry in string theory, or supersymmetric gauge theories in general), ``sigma models (?)'', ``Higgs sector beyond the \gls{sm}'', ``supergravity'', ``Higgs boson'', ``supersymmetric particles'', ``flavor physics''. The meaning of the ``sigma models'' context is unclear, although it comprises most occurrences of terms relating to superspaces and superfields. These concepts are directly tied to supersymmetry. They arise from the abstract extension of space by introducing extra anti-commuting coordinates. That supersymmetry spans across distinct topics constitutes evidence for the diversity of its uses. It is also notable that several of these topics are in fact dominated by supersymmetry (``supersymmetric theories'', ``supergravity'' and ``supersymmetric particles''). This stresses the importance of supersymmetry in the high-energy physics literature.

Moreover, although all these words (``supersymmetry'', ``supersymmetric'' and ``susy'') should refer to the same concept, we find that they are in fact related to different topics: ``supersymmetry'' seems to encompass more theoretical aspects of supersymmmetry (e.g. supergravity) while ``susy'' is more likely to occur in relation to supersymmetric particles (phenomenological supersymmetry). In fact, we find that 60\% of papers mentioning ``supersymmetry'' belong to theory (versus $\sim$40\% to phenomenology) while only 30\% of papers mentioning ``susy'' in their abstract belong to ``theory'' (versus 70\% to phenomenology).

\begin{figure*}
    \centering
    \hspace*{-2em}\includegraphics{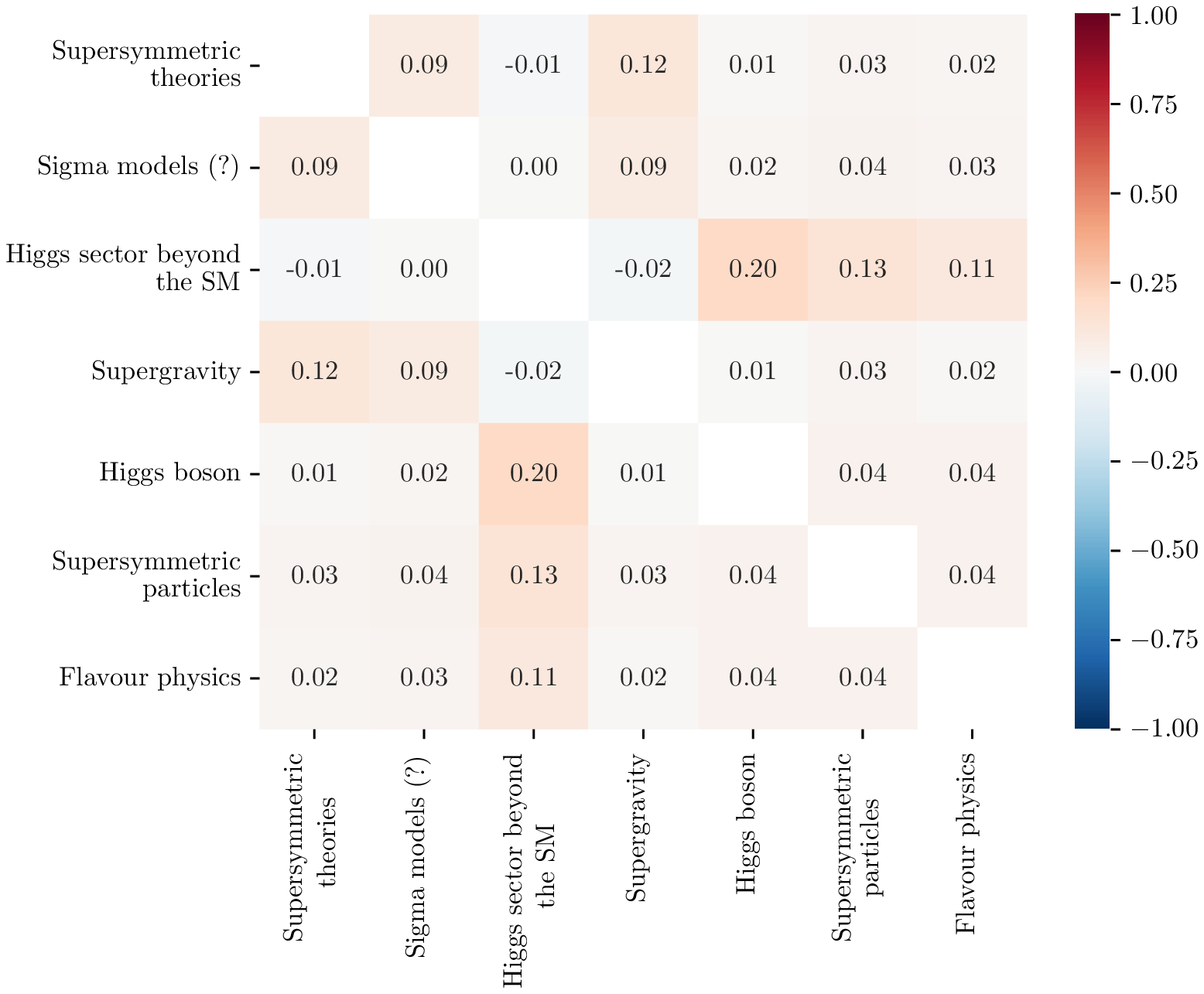}
    \caption{\textbf{Correlation between the topics most associated to supersymmetry}. The Pearson correlation is comprised between -1 (perfect anti-correlation) and 1 (perfect correlation). A correlation close to 0 means that a pair of topic is partially independent, i.e. that they can arise or not in variable proportions in a paper.}
    \label{fig:susy_correlations}
\end{figure*}

That these topics are at least partially independent can be assessed by inspecting the covariance matrix $\Sigma$ of the Correlated Topic Model from which they were derived. We therefore compute the correlation matrix\footnote{The Pearson correlation coefficients $R_{ij}$ can be deduced directly from the covariance matrix $\Sigma$ of the CTM model -- cf. equation \eqref{eq:ctm_prior} -- according to $R_{ij} = \dfrac{\Sigma_{ij}}{\sqrt{\Sigma_{ii}\Sigma_{jj}}}$} between the seven topics most commonly associated with supersymmetry; the results can be found in Figure \ref{fig:susy_correlations}. Overall the correlations are close to 0, which suggests that these topics are rather independent, with a few exceptions. In particular, pairs of topics that belong to the same kind (theoretical or phenomenological) are moderately correlated; pairs of topics that are directly tied to supersymmetry (e.g., supergravity and phenomenological supersymmetry) but of different nature (in this case, theoretical and phenomenological, respectively) are less correlated. Further visual evidence is provided in Figure \ref{fig:tsne} (Appendix \ref{appendix:topics_categories}).

From these results, one can see that supersymmetry is itself a diverse concept. It arises in a variety of partially independent contexts. In particular, theoretical and phenomenological aspects of supersymmetry are quite independent. How have these different aspects of supersymmetry evolved after the negative results of the searches for supersymmetric particles at the \gls{lhc}?

In order to address this question, we evaluate the evolution of supersymmetry research in HEP since the first results of the LHC (2011) until today. For that, similarly to \citet{Hall2008}, we assess the relative importance $\hat{\theta}_{z,y}$ of each topic $z$ for every year $y$ from 2011 to 2019:

\begin{equation}
P(z|y) = \dfrac{1}{D_y}\sum_{d\in y} P(z|d)   
\end{equation}

\begin{figure*}
    \centering
    \includegraphics{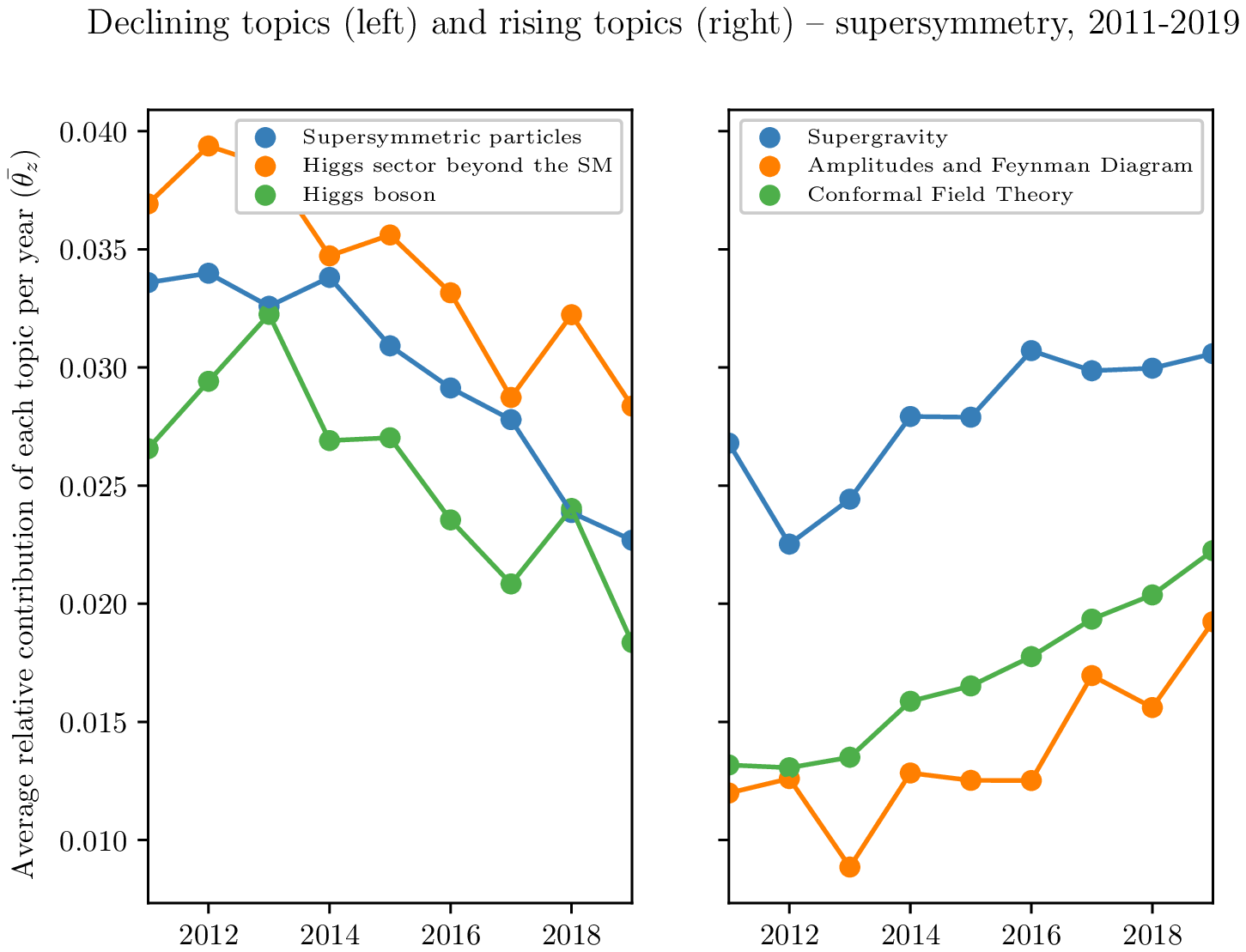}
    \caption{\textbf{Declining and rising topics among those that mention supersymmetry since the first results of the LHC (2011-2019).} On the left, the three topics that are declining the fastest ``Supersymmetric particles'', ``Higgs sector beyond the SM'' and ``Higgs boson''. On the right, the three fastest rising topics are ``Supergravity'', ``Amplitudes and Feynman Diagrams'', ``Conformal Field Theory''.}
    \label{fig:hot_cold_topics_hep_2011_2019_susy}
\end{figure*}

Where $D_y$ is the amount of articles first submitted in year $y$. We then selected the three topics with the highest increase (rising topics) and decrease (declining topics) in magnitude over this period. For that, $P(z|y)$ was fitted to a linear time trend ($P(z|y) = a_zy+b_z$), discarding topics for which the correlation was not significant (i.e. $R=0$ is excluded from the 99\% CI). Then, the topics were sorted according to the best fit value of $a_z$, the rate of increase of its magnitude per year (similarly to what was done in \citealt{Griffiths2004}). We apply the procedure to  all papers mentioning at least one of the words ``supersymmetric'', ``supersymmetry'' or ``susy'' in their title or abstract in the years following the start of the \gls{lhc}. The results are shown in Figure  \ref{fig:hot_cold_topics_hep_2011_2019_susy}.

According to these results, the most rapidly declining topics among articles that mention supersymmetry are Higgs-sector related topics and phenomenological supersymmetry, i.e. phenomenological aspects of supersymmetry. By contrast, two of the (relatively) increasingly active topics are very theoretical (in particular, Supergravity and Conformal Field Theory). 
In order to understand these dynamics, it is therefore necessary to distinguish theoretical supersymmetry from phenomenological supersymmetry. As physicist Mikhail Shifman argued in an early assessment of the first results of the \gls{lhc} in 2012, 

\begin{quote}
    [Theoretical supersymmetry] is an example of a complete success story. I use the word ‘theoretical’ to differentiate from ‘phenomenological’ supersymmetry [\dots] which [\dots] at the moment has a rather murky status. Theoretical supersymmetry proved to be a powerful tool with which to deal with quantum field theory, especially at strong coupling, a regime which was considered intractable for decades[\dots]. Progress in this line of research [\dots] is absolutely steady (\citealt[p.~6]{SHIFMAN2012})
\end{quote}



Shifman's assessment strikingly converges with the patterns that emerge from our analysis. Topic models reveal the plurality of supersymmetry in high-energy physics. They support that supersymmetry arises in different contexts, some theoretical and others phenomenological. They allowed us to demonstrate that these ``faces'' of supersymmetry have responded differently to the absence of evidence for supersymmetric particles at the \gls{lhc}. Indeed, although phenomenologists find supersymmetry to be much less valuable in the light of the most recent experimental findings, theorists may still rely on it for their own endeavor. This supports that cultures can ``trade'' certain concepts (according to Galison's terminology) while retaining much of their autonomy, including in their own appraisal of the usefulness of these concepts\footnote{``trading partners can hammer out a local coordination, despite vast global differences.'' \citep[p.~783]{galison1997image}}. 

In the next section, we investigate the contribution of supersymmetry to sustaining the trading zone between these theoretical traditions throughout time.

\subsection{Supersymmetry in the trading zone between theory and phenomenology}

\begin{figure}
    \centering
    \includegraphics{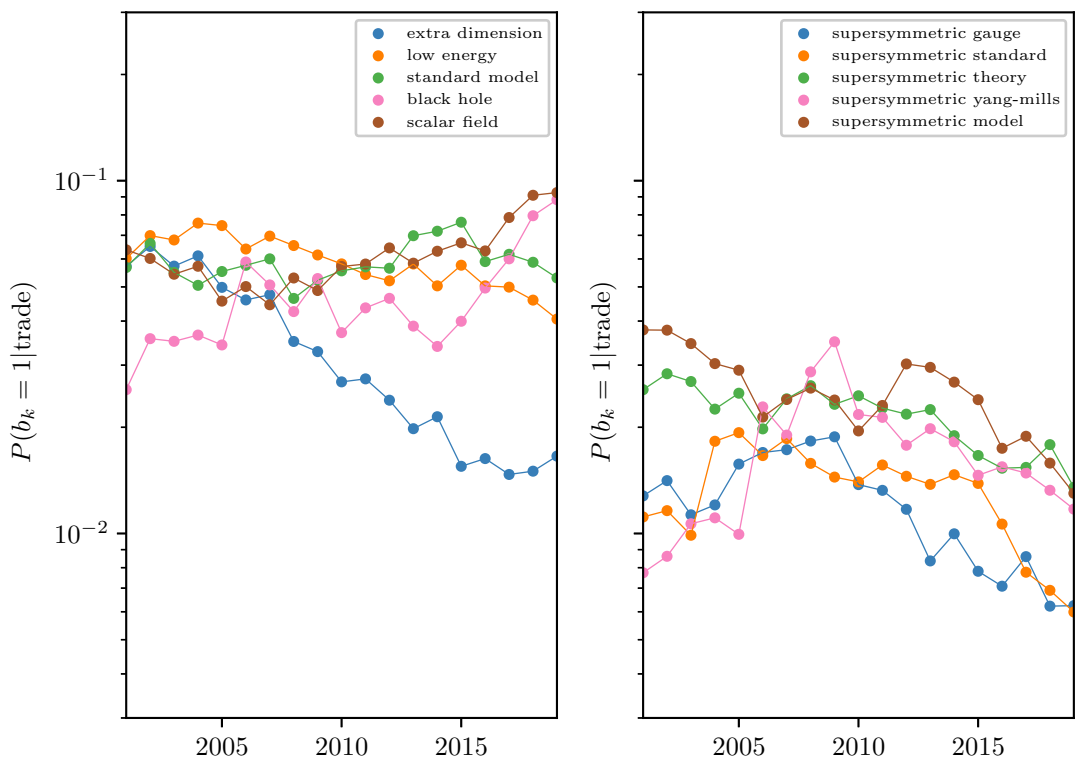}
    \caption{\textbf{Inside the trading zone: probability that certain keywords appear in the abstract of a theoretical paper involved in a trade (a phenomenological paper citing a theoretical paper).} To the left, the five keywords are those with the highest peak probability of occurrence; to the right, are the five keywords with the highest probability of occurrence among supersymmetry related keywords. Redundant keywords (which normalized pointwise mutual information with a more frequent keyword exceeds 0.9) are excluded.} 
    \label{fig:cross_citations_probs_th_ph}
\end{figure}

\begin{figure}
    \centering
    \includegraphics{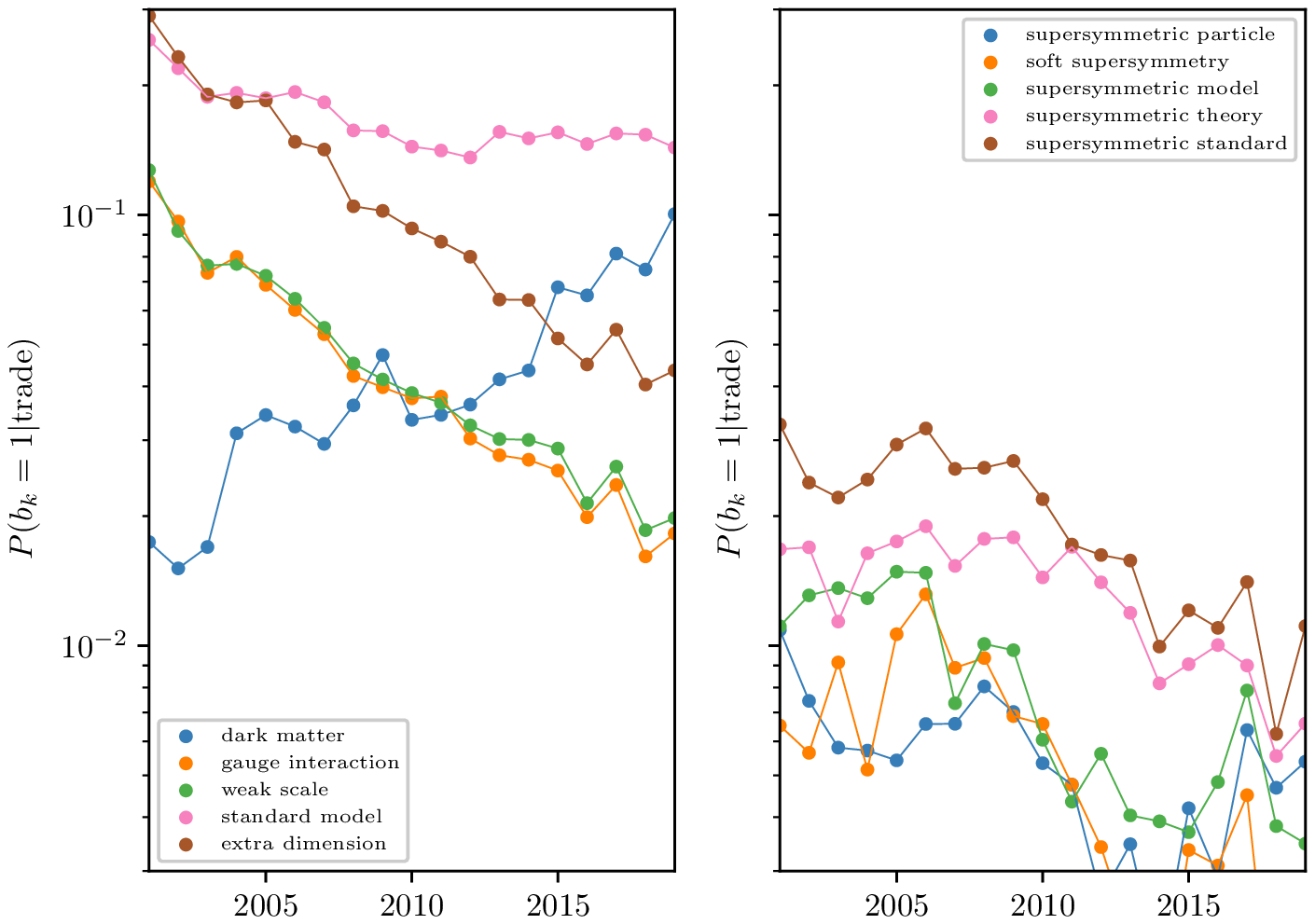}
    \caption{\textbf{Inside the trading zone: probability that certain keywords appear in the abstract of a phenomenological paper involved in a trade (a theoretical paper citing a phenomenological paper).} To the left, the five keywords are those with the highest peak probability of occurrence; to the right, are the five keywords with the highest probability of occurrence among supersymmetry related keywords. Redundant keywords (which normalized pointwise mutual information with a more frequent keyword exceeds 0.9) are excluded.} 
    \label{fig:cross_citations_probs_ph_th}
\end{figure}

Which concepts sustain trades within \gls{hep}? As proposed in Section \ref{section:method_trading_zone}, we measure the ability of certain concepts (keywords) to sustain trades through time in terms of the probability that each of these concepts occurs in citations across theory and phenomenology. The results are shown in Figures \ref{fig:cross_citations_probs_th_ph} and \ref{fig:cross_citations_probs_ph_th}.

Both these figures show the probability of occurrence of the five most common keywords (left side) and the five most common supersymmetry-related keywords (right side) involved in trades across these subcultures (excluding redundant keywords). Figure \ref{fig:cross_citations_probs_th_ph} shows those probabilities for trades where phenomenological papers draw from theoretical papers. Three main trends are revealed: the fall of trades involving extra-dimensions (hypothesized spatial dimensions beyond the 4 space-time dimensions for which we have direct evidence); the increase in trades involving black-holes; and, directly relevant to our third claim, the decline of trades involving supersymmetry, despite a short increase after the start of the \gls{lhc} in 2010. Interestingly, in the early 2000s, ``supersymmetric model[s]'' had a trade-ability on par with that of the keywords most involved in these trades. Moreover, ``low energy'' appears to be one of the most frequent keywords in phenomenological imports of theoretical papers, which makes sense since the low-energy limit of theories of, say, strings and quantum gravity is what matters most from a phenomenological standpoint (it is what can be observed). Turning to Figure \ref{fig:cross_citations_probs_ph_th} -- trades involving theoretical references -- , we get an even more striking picture of the demise of ``extra dimensions'', which were involved in about 30\% of the trades in 2001 and went down to 5\% only. Similarly, ``weak-scale''  which refers to the domain of phenomena targeted by the \gls{lhc}, have become much less frequent in the ``trading zone'' (from $\sim$10\% of trades to $\sim$2\%). This suggests that phenomenological models dedicated to this domain of phenomena have become much less useful to the ``theoretical'' subculture over time. On the other hand, ``dark matter''\footnote{Dark matter refers to the observation that a significant fraction of the mass of the universe is currently unexplained.} is increasingly common in the phenomenological papers theorists draw from. This suggests dark matter is deemed valuable for the theoretical enterprise as well. This figure also confirms the overall decline of supersymmetry in the trading zone, thus providing further support to our third claim:  supersymmetry does not connect developments from the theoretical programs to progress in the phenomenological program as much as it did prior to the \gls{lhc}.

\section{\label{section:discussion}Discussion}

\subsection{Conceptually informed methods in quantitative science studies}

Before exploring the implications of this case-study, we want to emphasize that Galison's conceptual framework has been a fruitful guide for our quantitative approach. The linguistic component of his notion of subculture led us to build a bag-of-words model for measuring the extent of the divide between two theoretical cultures, and for unveiling the concepts that are specific to these cultures as well as their methodological and ontological differences. The social autonomy of these subcultures, too, can be readily quantified from authorship data. Furthermore, the notion of trading zone invited us to explore citations quantitatively (as a proxy of scientific ``trades'') while devising ways to determine their ``location'' in the semantic space. We also found that topic models can reveal the plurality of contexts in which a concept may arise, and how the dynamics of these contexts compare throughout time. Although we have applied our topic model approach to supersymmetry, in principle it can be applied to any kind of ``boundary object'', understood in the broad sense of a shared notion that allows some coordination to be achieved while preserving the distinctness of the scientific cultures at play. In the end, these methods illuminated our study of supersymmetry in high-energy physics, and provided further grounds for Galison's claim that unity is a contingent matter.

\subsection{Unity challenged?}

The two theoretical subcultures we have distinguished -- pure ``theory'' and phenomenology -- no longer seem to value supersymmetry equally. Supersymmetry indeed fails to provide equally satisfying solutions to the heterogeneous commitments of \gls{hep} physicists, which poses a challenge to the unity of the field. Indeed, the example of supersymmetry shows that what drives theoretical progress may not drive phenomenological progress -- in contrast with the expectations of the community regarding supersymmetry prior to the \gls{lhc} as surveyed by \citealt{Mttig2019,Mttig2020a} -- and developments in these subcultures may become quite orthogonal.

Of course, supersymmetry is not the only channel of coordination between the theoretical and phenomenological cultures in their search for ``new physics''. Another channel, for instance, has been the notion of extra dimensions (see Figures \ref{fig:cross_citations_probs_th_ph} and \ref{fig:cross_citations_probs_ph_th}), which dominated trades in the early 2000s to an extent we did not expect before conducting this analysis. Extra-dimensions are required by string theory, but they are also subject to trades with phenomenologists interested in their observable consequences. However, no evidence for extra-dimensions was found at the \gls{lhc}. This further supports that the goals that drive theoretical research programs such as string theory (like the search for a quantum description of gravity) may not serve the phenomenologists' agenda so well.

Eventually, the \gls{lhc} provided ``a test of the unity of physics''\footnote{ \citet[p.~29]{Wilson1986} (cited in \citealt[p.~292]{Cat1998}) used this expression in reference to the now aborted Super-Conducting Supercollider.}, and its verdict was ruthless. In the future, will the field strive to regain unity (possibly to the detriment of certain research programs), or will the socially entrenched divergences between these ``cultures'' of high-energy physics prevail?  We may assume that the challenge is merely transitory, and that theorists will eventually move to other theories which will be more successful from an empirical or phenomenological standpoint. However, the divergence between these theoretical cultures has become axiological \citep{Camilleri2015,Laudan1984}, in the sense that they prioritize different epistemic goals\footnote{ \citet{Laudan1984} refers to disagreements in the goals of scientific inquiry as \textit{axiological} disagreements. \citealt{Camilleri2015}, for instance, have argued that certain controversies around string theory could be understood in terms as an instance of axiological disagreement.}; and this divergence may persist as long as their differences in aims persist; as \citeauthor{galison1997image} puts it, ``there is no teleological drive towards ever-greater cohesion'', and ``fields previously bound [may] fall apart'' \citep[p.~805]{galison1997image}. As illustrated in Figure \ref{fig:mssm}, the aims of the theorists is to achieve the unification of the fundamental forces and a coherent theory of quantum gravity. By contrast, the aim of phenomenologists is to guide the experiment towards promising directions where evidence of ``new physics'' may be found. Both these aims may seem well-founded; however, there is no reason to expect that a simultaneous solution can be worked out. The apparent failure of supersymmetry to provide such a simultaneous solution does not undermine by itself the relevance of the ``theorists''' aims, nor does it undermine the methodology they deploy for addressing their goals (e.g. their trust in certain theoretical constraints, cf. \citealt{Galison1995}). It does, however, challenge the belief that such methods can provide grounds for progress to the field \textit{as a whole}; indeed, unification and quantum gravity might eventually not provide much reliable guidance to the experimental side. Conversely, it can very well be that the details of the theory ``at high energy'', where quantum effects matter to gravity, cannot be extrapolated from our knowledge of the low-energy theory, i.e. the one that we can probe in our experiments. As a result, \citet{Dawid2013} argues for recourse to meta-empirical assessment of theories in theoretical physics, given that empirical input under-determines the directions of potential progress in quantum gravity. Disagreements in the aims of a scientific enterprise may not always be resolved on purely epistemic ground, and a resolution, provided it occurs, may involve some sort of negotiation instead.  As long as theorists believe in the feasibility of their aims, they may pursue these aims even if it further isolates them from other cultures\footnote{More drastically, \citet{Cao1993} expressed the view that the theories at different energy scales (i.e. corresponding to different ranges of phenomena) are irreducible, and they argued for a ``pluralist view of possible theoretical ontologies'' while challenging the possibility of achieving a ``ultimate stable theory  of everything'' (p.~69--71). According to this view, the plurality of ontologies in physics is not an accident but the result of partially disconnected ``phenomenological domains'' through which knowledge cannot be deduced from one another. For a criticism of this view, see \citealt{Rivat2020}.}. Alternatively, they could decide that the schism should be resolved; as Galison puts it, distinct scientific cultures ``can [\dots] understand that the continuation of exchange is a prerequisite to the survival of the larger culture of which they are part'' \citep[p.~803]{galison1997image}.

\subsection{Trading zones as a mean to sustain diversity}

More generally, the example of \gls{hep} and supersymmetry demonstrates how disunity can be endogenously produced in the fabric of science. Even initially tightly bound scientific cultures can diverge into quite distinct and autonomous programs, with different ontologies, methodologies and aims, as new domains of inquiry open up (e.g. quantum gravity) and warrant new modes of knowing. The extent of the coordination between disciplines will in general depend on epistemic factors (depending on how fruitful certain ``trades'' turn out to be), but also on non-epistemic factors: for instance, it may depend on the institutional setting, or whether such exchanges are incentivized or ``coerced'' \citep{Collins2010}.

Paradoxically, it can be noted that trading zones can stabilize the heterogeneity of cultures within a field, by sustaining the practitioners' beliefs that, in spite of the large differences in what they are doing, their respective efforts somehow support each other. If that is the case, there is no perceived need for a profound re-alignment of their respective practice. Trading zones can contribute, therefore, to a mutual process of legitimization of heterogeneous scientific practices, which is not necessarily tantamount to further ontological unity. In order to further emphasize that, it is useful to come back to the example of \gls{hep}, and most particularly that of string theory, a highly theoretical research program driven by the pursuit of a consistent theory of quantum gravity. String theorists such as Matt Strassler have argued that even if string theory did not directly provide testable predictions to phenomenologists and experimentalists, it generated mathematical tools that could be useful to their practice, e.g. for predicting the behavior of quark-gluon plasma \citep{Ritson2021}. Consequently phenomenologists may have a low appraisal of string theory in terms of its ability to generate models for testing its assumptions about nature, while still recognizing the usefulness of what string theorists do for them, since some of their work is effectively ``applicable''. As \citet{Ritson2015} put it, ``if string theory has proved so useful for branches of physics whose scientific status is not in question, it can be argued it forms a legitimate part of physics''. Supersymmetry itself may be experiencing the same fate, considering that ``supersymmetry as a tool for exploring gauge dynamics at strong coupling [\dots] is taking precedence over phenomenology'' (\citealt[p.~7--8]{Shifman2020}). 
Such trades do support the usefulness of the theoretical program to other endeavors, without necessarily implying further integration of the subcultures of \gls{hep} (ontological unity); just like successful interdisciplinary work does not necessarily amount to further integration of disciplines \citep{GrneYanoff2016}. 

\subsection{Limitations and future work}

Before concluding we would like to hint at several directions for future work that could overcome certain limitations of the present methodology and further inform the question of the disunity of science.

First, none of our semantic methods distinguished between different kinds of words, i.e. which words refer to, say, methods (such as computation techniques) rather than entities  (e.g. strings, particles, etc.). It would be interesting to evaluate to what extent the coordination between theoretical cultures involves ontological or mere methodological trades, depending on whether the constructs of high-theory are referred to as the proper description of nature or as mere mathematical tools, and how this may have changed throughout time. This might uncover evidence for a shift from an ontological to a more methodological coordination between the subcultures of high-energy physics, as the arguments for supersymmetry and string theory as ``tools'' rather than accurate accounts of the natural world suggest.

Another direction of future work involves the topic model approach. Although the topic model used in this work yielded seemingly acceptable results overall, some topics were difficult to interpret. In that respect, we made several improvements compared to previous works, by training the model on not just single words but also n-grams matching specific and presumably semantically informative syntactic patterns and by informing our interpretation of topics using correlations with a standard classification (rather than the top-words only). Yet, further improvements could be made. First, vocabulary selection could be enhanced by a better handling of mathematical expressions, for instance by parsing LaTeX formulas. The NLTK library picked up some of these expressions, and since they captured some information about the documents, we did not exclude them from the vocabulary; however, this way of proceeding does not preserve the underlying mathematical structure, although it may be valuable to distinguish references to, say, specific particles, or certain symmetry groups, based on their mathematical notations. We may also want the model to learn to discard uninformative words such as ``result'', ``parameter'', ``model'', etc.. In our case, we found such vague words to be clustered in three topics that we labeled as ``jargon'' which correlated very poorly with the standard classification (see Tables \ref{table:top_words} and \ref{table:full_topics_pacs_pmi}), but they should ideally not emerge as distinct topics on \textit{par} with more meaningful topics. To this end, we may want to build on \citealt{syntax_topic_model}, which provides a model that is able to distinguish between ``semantic'' and purely ``syntactic'' clusters of words without  prior knowledge of the language. A more critical limitation of topic models pertains to the challenge of hyper-parameters' tuning, considering it is unclear which performance metric should be maximized in the process. Although we proposed a procedure for choosing these parameters that accounts for known limitations to the reliability of perplexity or topic coherence metrics, non-parametric methods may provide a better answer to this fundamental issue \citep{Gerlach2018}.

Finally, the historical scope of our analysis was limited by our database. In particular, we were only able to analyze the theory/phenomenology divide over a restricted time range (1980--2020), and we could not reveal how such a divide has historically emerged. By contrast, Galison has proposed a number of explanations for the earlier decoupling of theory and experiment, such as increased specialization and the increased time-scales of experiments \citep[p.~138]{galison1987how}.

\printglossary[type=\acronymtype,title=List of abbreviations]


\begin{acknowledgements}
We would like to thank Olivier Darrigol for his extremely insightful comments regarding the concept of trading zones; Arianna Borrelli for commenting on the work that led to this publication; Thomas Heinze and  Radin Dardashti for their comments and continuous support; Alexander Blum for his consideration and his suggestions; and Elizabeth Zanghi for her corrections. Finally, we thank all reviewers for their in-depth comments.
\end{acknowledgements}

\section*{Funding}

The authors acknowledge support from the Open Access Publication Fund of the University of Wuppertal.

\section*{Conflicts of interest}

The authors declare that they have no conflicts of interest.

\section*{Data availability}

All the data and code used to derive the results of this paper can be accessed from the following repository using the DataLad software \citep{datalad_paper}: \url{https://gin.g-node.org/lucasgautheron/trading_zones_material}. Upon publication, the repository will be archived and a DOI will be included in the present paper.

\printbibliography

\appendix

\section{Appendices}
\label{section:appendix}

\subsection{Data collection}
\label{appendix:collection}

Our goal was to collect the whole HEP literature from 1980 to 2020 from the public Inspire HEP API \citep{InspireAPI}. For that, we collected metadata for all articles through automated search requests, category per category, and year per year. This strategy was intended to abide with the limitations of the API, in terms of matching entries per search request. However, it appeared that many articles in years 1990 to 1995 were not categorized, and therefore our collection strategy missed many HEP articles from this period. In order to recover these articles, we gathered all articles that were referenced in publications collected through the first batch but which were missing. This methods fails to recover articles that were not cited in any article from the first batch. More importantly, 
the lack of categories means that selecting all HEP papers during the problematic time period will require unlabeled articles to be manually or automatically classified. Although there are ways to circumvent these issues and to assess their potential implications, we have decided to narrow down several analyses to years 2001 onwards in the present work.


\subsection{\label{appendix:stability}Text-classifier performance stability}

\begin{figure}
    \centering
    \includegraphics{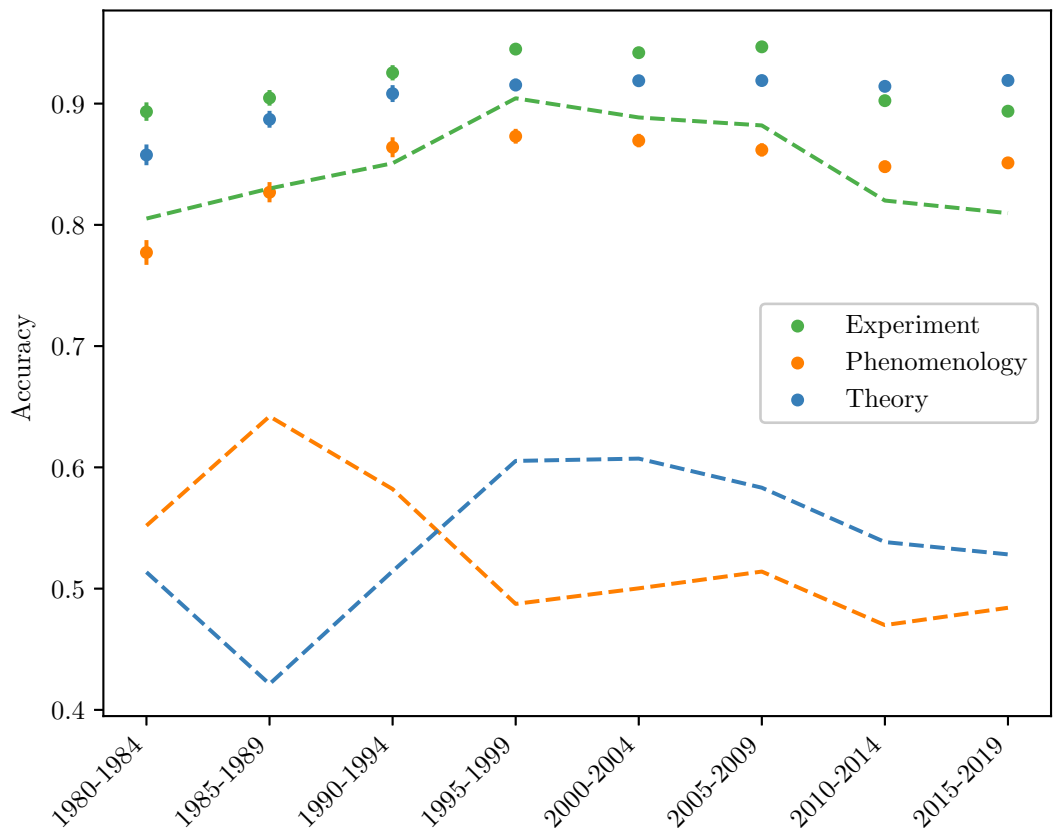}
    \caption{Accuracy of the text-classifier from Section \ref{section:method_subcultures} as a function of the papers' years of publication. Error-bars represent the 95\% confidence interval. Dashed lines show the accuracy of the baseline model (which may vary only due to variations in the frequency of each category, since the baseline model always predicts the most common class). The accuracy is roughly constant across time for each of the three categories, despite significant variations in the frequency of each class.}
    \label{fig:stability}
\end{figure}

The categories (\texttt{Theory-HEP}, \texttt{Phenomenology-HEP} and \texttt{Experiment-HEP}) that we trained our classifier (\ref{section:method_subcultures}) to predict have been assigned in different ways in the Inspire HEP database. Although a majority were categorized based on arXiv's classification system, some papers were not, especially those published before arXix was introduced (in the early 1990s). It might seem unclear whether these classification procedures are consistent and revealing of distinct underlying cultures. In order to demonstrate that it is the case, in Figure \ref{fig:stability}, we show that the performance of the text-classifier is nonetheless roughly stable throughout the period considered (1980--2020). To this end, we subdivide this time-range in bins of five years and perform k-fold cross-validation using each five year bin for the validation set (and the papers from the other bins for the training set). Accuracy remains high and approximately stable over the years 1980 to 2020; therefore, these various classification procedures, and the underlying identity of each of these subcultures, must be rather consistent over this period.

\subsection{Topic model}

\subsubsection{\label{appendix:data_selection}Data and vocabulary selection}

The model is trained on $N=120,000$ articles randomly sampled from those in the 1980-2020 period that belong to any of the categories \texttt{Theory-HEP}, \texttt{Phenomenology-HEP}, \texttt{Experiment-HEP}, and \texttt{Lattice}. Titles and abstracts of each papers are concatenated in order to maximize the textual content used for training. Very short texts (less than 100 characters) are removed.

Before applying the model, we performed a number of pre-processing steps on the abstracts with the goal of maximizing the amount of useful information in the training data. This procedure, largely inspired from \citealt{omodei_tel-01097702} and implemented with the use of the NLTK library \citep{nltk}, is as follows:

\begin{itemize}
    \item Tokens (words separated by punctuation or spaces) are extracted from the text and transformed to lower-case.
    \item All single nouns and adjectives are retrieved from these tokens.
    \item We also retrieve all n-grams that match specific syntactic patterns (e.g. ``adjective+noun+noun'', such as ``supersymmetric standard model'', ``effective field theory'').
    \item Single words are lemmatized, i.e. they are normalized to their root (e.g. ``symmetries'' becomes ``symmetry'').
    \item Words and expressions that occur less than 20 times are removed.
\end{itemize}

First, these steps allow us to reduce noise by removing words that convey little to no information about the topics of the articles (such as stop words). Second, extracting n-grams that matching certain syntactic patterns allows us to preserve some information about the relative position of words within the abstracts -- which CTM do not do otherwise -- while taking advantage of our prior knowledge of the documents' language. For instance, the word ``dark'' may convey different meanings depending on whether it occurs immediately before the word ``matter'', or, alternatively, ``energy''; similarly, the occurrence of the expression ``dark matter'' in a text conveys more information than the simultaneous occurrence of ``dark'' and ``matter'' without more knowledge about their relative position.

As a result of this procedure, the vocabulary contains $V=$ 18,658 ``words'', with 58 words per article on average.

\subsubsection{\label{appendix:hyper_parameter}Hyper-parameters}

The implementation of the CTM by Tomotopy \citep{tomotopy} has three hyper-parameters: the amount of topics $k$, an $\vec{\alpha}$ parameter that controls the sparsity of the document-topic distribution ($\theta_{d,i}$), and a $\vec{\eta}$ parameter that controls the sparsity of the topic-word distribution (the vocabulary associated to each topic).
For choosing the amount of topics $k$, we considered three values that seemed acceptable in terms of interpretability and compliance with the values from the literature: 50, 75 and 100.
We assumed $\vec{\alpha}$ and $\vec{\eta}$ to be symmetric, i.e. $\alpha_1 = \alpha_k = \alpha$ and $\eta_1 = ... = \eta_V = \eta$\footnote{This is common in the literature, but this choice is disputable, cf. \citealt{Wallach2009}. One implication of symmetric priors is that topics must have comparable probabilities. This also has an impact on the meaning of topics.}. We considered  $\alpha \in \{10^{-2},10^{-1},1\}$ and $\eta \in \{10^{-3},10^{-2},10^{-1}\}$, according to values encountered in the literature.
We then trained the model for each triplet of $k$, $\alpha$ and $\eta$ among the candidate values. We rejected all triplets that led to significant overfitting, by comparing the perplexity\footnote{Perplexity is the exponential of the average log-likelihood per word, cf. \citealt{Blei2003}. It measures the improbability of a corpus according to a given model.} obtained for the training corpus and that obtained by applying the trained model to a validation set of abstracts unseen during training.
Although \citet{Chang2009} have shown that perplexity could be negatively correlated to human judgments about the interpretability of the topics recovered by topic models, we believe it is a suitable metric to discard models that fail to capture meaningful regularities in the data, which is the case of models that show overfitting. Among the remaining models, we then selected the two models with the highest normalized pointwise mutual information coherence, a coherence metric frequently used to assess the consistency of topic models \citep{hoyle2021is}. Topic coherence metrics in general, as stressed by \citeauthor{hoyle2021is}, are not very strongly correlated with human judgments about the quality of a model; however, we believe they may be useful to discard certain models in order to limit the amount of those that should be inspected manually (since manual inspection is time-consuming and quite subjective). We finally inspect manually the two models with the highest coherence measure, and choose the one with $k=75$, $\alpha=0.1$ and $\eta=0.001$. Our preference for this model stemmed from the fact that it contained more topics than the other remaining model, and that these more numerous topics seemed reasonably consistent.

\subsubsection{Validation}
\label{appendix:validation}

Since the model infers document-topic distributions and topic-word distributions, we would like to assess the validity of these metrics, i.e. ``their ability to measure what they purportedly measure'' \citep[p.~3240]{Bannigan2009}. In order to simultaneously assess both measures, we designed the following protocol. First, we derived the \gls{pacs} categories $c$ that were the most correlated to each topic $z$ (this approach is in a sense comparable to that employed in \citealt{Griffiths2004}, who extracted the topics that were more strongly associated with PNAS categories). For that, we listed the categories $c$ that maximize the pointwise mutual information with each topic $z$ according to:

\begin{equation}
    \label{eq:pmi_expression}
    \mathrm{pmi}(z,c) = \log \dfrac{p(z|c)}{p(z)}
\end{equation}

Where $p(z)$ is the marginal probability of the topic $z$, and $p(z|c)$ the probability that a word in a document belongs to a topic $z$ given that the document was assigned the PACS category $c$. Thefore, $\mathrm{pmi}(z,c)$ measures the increase in probability of a given topic provided that a PACS category is present. The 5 categories most correlated to each topic are given in table \ref{table:full_topics_pacs_pmi}, which helped inform our choice for each tpic label, in complement to their top-words.

Then, we submitted the lists of PACS categories thus constitued to a human task derived from the methodology of \citet{Bennett2021}, as follows:

\begin{enumerate}
    \item We draw at random a topic $z_1$ with a probability equal to its marginal probability 
    \item We draw at random 5 PACS categories $c_1,...,c_5$ among the 10 most correlated to $z_1$, as described above.
    \item Then, we do any of the following, with equal probability $1/2$:
    \begin{enumerate}
        \item We draw at random another topic $z_2\neq z_1$ with probability $p(z_2)\over 1-p(z_1)$, and we pick at random 5 PACS categories $c_6,...,c_{10}$ among those most correlated with it.
        \item Alternatively, we draw  $c_6,...,c_{10}$ from the 5 remaining PACS categories most associated to $z_1$
    \end{enumerate}
    \item We submit $c_1, ..., c_5$ and $c_6, ..., c_{10}$ to an expert unaware of the model. The expert is asked to guess whether the two lists of 5 categories were drawn from one and same general topic, or whether they were drawn from two separate topics.
    \item The procedure is repeated a certain amount of times. The final score is the fraction of correct responses.
\end{enumerate}

The rationale for this method is that good scores should only be achievable provided the topics are rather coherent, and that the document-topic distributions $\theta_{d,i}$ are reasonably accurate. The final average score is 0.74 for 100 guesses from two HEP PhD students, which is significantly better than a random baseline (0.5). This shows that, to some extent, the topic distributions derived for each article correlate with \gls{pacs} categories that are rather coherent with each other.

\subsubsection{Topics}


\fontsize{6}{7}\selectfont\input{Table4}\normalsize
\fontsize{6}{7}\selectfont\input{Table5}\normalsize

\subsubsection{\label{appendix:topics_categories}Topics and their correlation with categories}

\begin{figure}
\centering
\includegraphics{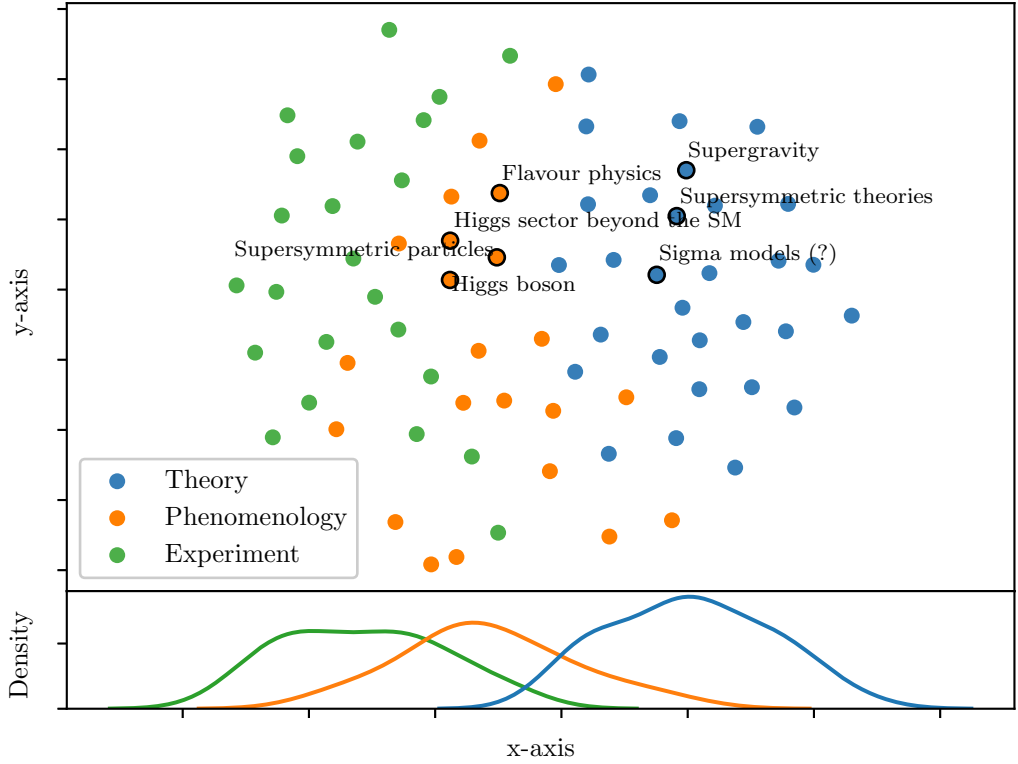}
\caption{\label{fig:tsne}Semantic map extracted from the topic model, after applying a t-SNE transformation. Each dot represents a topic. Each topic is assigned the category, among theory, phenomenology and experiment, that is most associated with it. Correlated topics appear closer to each other. For each category, the density of topics along the x-axis is shown in the lower plot.}
\end{figure}

Below, we evaluate how topics compare with the classification of the literature. For that, we generated a 2D representation of the semantic space by applying a t-SNE transformation \citep{Maaten2008} on the distance matrix $1-R_{ij}$, where $R_{ij}$ is the correlation matrix for the 75 topics from the CTM. The t-SNE transformation aims to reduce dimensionality (from 75 to 2) while preserving distances, such that highly correlated topics should appear close to each other on the resulting 2D map. We then colored each topic according to the category (among theory, phenomenology and experiment) that has the strongest association (normalized pointwise mutual information) with this topic. The graph was then rotated such that the x-axis would explain most of the variance in these three categories. Topics related to supersymmetry were emphasized and labeled. The resulting map is shown in Figure \ref{fig:tsne}.

Although the t-SNE transformation does not yield very stable results, it generally appears (as in this figure) that topics most associated with a given category (e.g. theory) appear closer to each other, such that these three categories explain part of the variance in the semantic space. Second, in this representation, the distinction between phenomenological supersymmetry and theoretical supersymmetry is supported by the emergence of two separate clusters of supersymmetry-related topics.

\subsection{\label{appendix:phenomenology_centrality}Validity of the citation network for exploring the trading zone}

Below, we support the relevance of the citation network as a means of exploring trading zones between scientific cultures by showing it can be used to recover known facts, in particular i) that theory and experiment in HEP do not communicate directly and ii) that phenomenology channels most exchanges across them.

We build a citation network where each node is one paper of the literature and the edge between nodes $x$ and $y$ is assigned a weight $w_{x,y}=1$ if $x$ cites $y$ and 0 otherwise. From this we can define the amount of citations of papers from the category $i$ to a paper from the category $j$ as: 

\begin{equation}
    \label{eq:cite_matrix}
    n_{ij} = \sum_{x\in i, y\in j} \dfrac{w_{xy}}{(\sum_c \mathds{1}_c(x))(\sum_c \mathds{1}_c(y))} 
\end{equation}

\begin{figure*}
    \centering
    \includegraphics{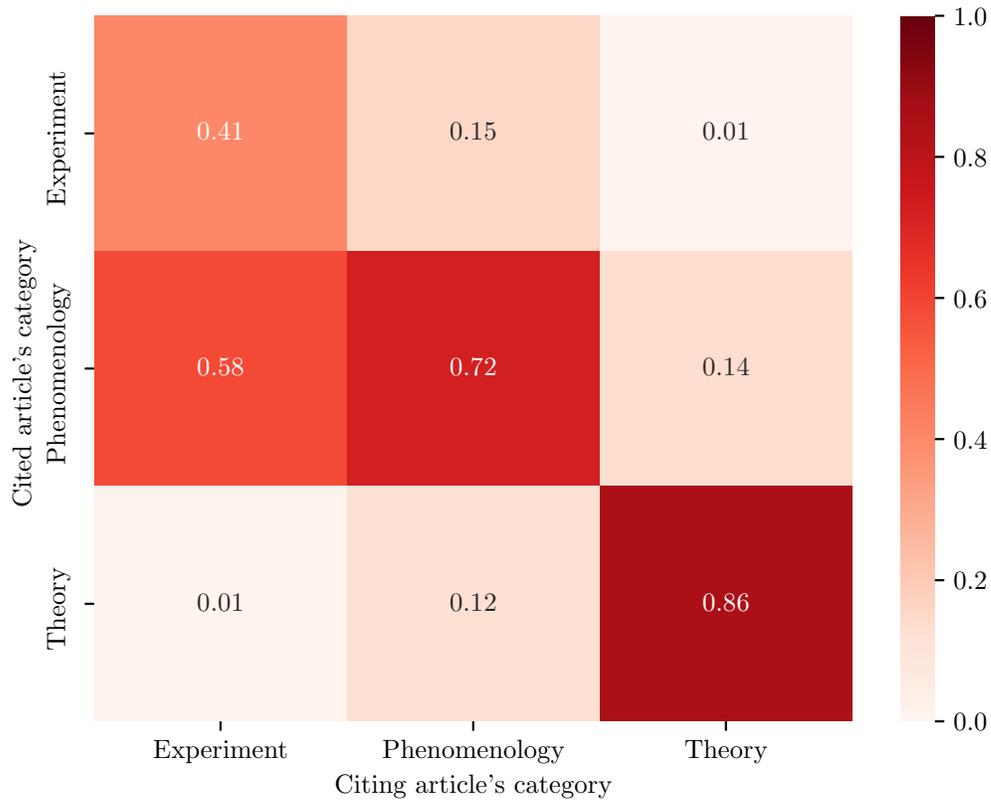}
    \caption{\textbf{Origin of the references (citations) in the \gls{hep} literature}
    Each matrix element $\tilde{n}_{ij}$ represents the fraction of references from the x-axis category (columns) that target papers from the y-axis category (lines). For instance, 41\% of references in experimental papers refer to experimental papers. 15\% of citations from phenomenological papers refer to experimental papers. If these categories were completely hermetic, the matrix would equal the identity matrix, which is not the case.}
    \label{fig:cites_matrix}
\end{figure*}

Where $\mathds{1}_c(x)=1$ if $x$ belongs to $c \in \{$Experiment, Phenomenology, Theory$\}$, and 0 otherwise. 
We then normalize $n_{ij}$ by the amount of citations \textit{from} category $i$, thus yielding the normalized matrix $\tilde{n}_{ij}$. By construction, $0\leq \tilde{n}_{ij}\leq 1$ is the effective fraction of references from papers of category $i$ to papers of category $j$. The matrix is built from the citation network between 2001 and 2019. We then verify that $\tilde{n}_{ii}$ is high (papers mostly cite papers from the same category); and that for cross-culture citations ($i\neq j$), $\tilde{n}_{ij} \ll 1$ unless $i$ or $j$ is ``phenomenology''; i.e., ``trading zones'' in the field occur around phenomenology. Evaluating the fraction of citations from papers of a category $i$ that target papers from a category $j$ yields the matrix in Figure \ref{fig:cites_matrix}. In this matrix, borrowing the trade metaphor from \citet{Yan2013}, non-diagonal elements represent ``imports'' (references to publications from other subcultures) and diagonal elements measure the ``self-dependence'' of each subculture. The results confirm that most citations occur within categories, emphasizing the relative autonomy of each of these subcultures including phenomenology -- it is less obvious for experimental papers, which are much more scarce then the others, and cannot cite themselves as much. Moreover the results confirm that most trades involve phenomenology: cross-citations between purely theoretical and experimental papers are very rare ($\sim$1\% of their references). Overall, ``theory'' is highly self-reliant.

\end{document}

%% file: acronyms_en.tex
\newacronym{doe}{DoE}{Department of Energy}
\newacronym{nsf}{NSF}{National Science Foundation}
\newacronym{oatd}{OATD}{Open Access Theses and Dissertations}
\newacronym{orcid}{ORCID}{Open Researcher and Contributor ID}

\newacronym{lep}{LEP}{Large Electron Positron Collider}
\newacronym{ssc}{SSC}{Superconducting Super Collider}
\newacronym{lhc}{LHC}{Large Hadron Collider}
\newacronym{fcc}{FCC}{Future Circular Collider}
\newacronym{eht}{EHT}{Event Horizon Telescope}
\newacronym{wos}{WoS}{Web of Science}

\newacronym{desy}{DESY}{\textit{Deutsches Elektronen-Synchrotron}}
\newacronym{slac}{SLAC}{Stanford Linear Accelerator Center}
\newacronym{cern}{CERN}{Conseil Européen de Recherche Nucléaire}

\newacronym{mssm}{MSSM}{Minimal Supersymmetric Standard Model}
\newacronym{pmssm}{pMSSM}{Phenomenological Minimal Supersymmetric Standard Model}
\newacronym{cmssm}{cMSSM}{constrainted Minimal Supersymmetric Standard Model}

\newacronym{aip}{AIP}{American Institute of Physics}
\newacronym{pacs}{PACS}{Physics and Astronomy Classification Scheme\textregistered}

\newacronym{sm}{SM}{Standard Model of Particle Physics}
\newacronym{ms}{MS}{Modèle Standard de la physique des particules}
\newacronym{hep}{HEP}{High-Energy Physics}
\newacronym{bsm}{BSM}{Beyond the Standard Model}

\newacronym{wimp}{WIMPs}{Weakly Interacting Massive Particles}

\newacronym{hi}{HI}{Historical Institutionalism}
\newacronym{cha}{CHA}{Comparative Historical Analysis}

%% file: Table2.tex
\begin{longtable}{p{7cm}|p{7cm}}
\caption{Vocabulary specific to phenomenology (left column) versus theory (right column).}
\label{table:specific_pheno_vocabulary_th_ph}\\
\toprule
                                                                                                                                                                                                                                                                                                                                                                                                                                                                                               Vocabulary specific to phenomenology &                                                                                                                                                                                                                                                                                                                                                                                                                                                                         Vocabulary specific to theory \\
\midrule
\endfirsthead
\caption[]{Vocabulary specific to phenomenology (left column) versus theory (right column).} \\
\toprule
                                                                                                                                                                                                                                                                                                                                                                                                                                                                                               Vocabulary specific to phenomenology &                                                                                                                                                                                                                                                                                                                                                                                                                                                                         Vocabulary specific to theory \\
\midrule
\endhead
\midrule
\multicolumn{2}{r}{{Continued on next page}} \\
\midrule
\endfoot

\bottomrule
\endlastfoot
quark, lhc, qcd, neutrino, experimental data, mssm, dark matter, extra dimension, parton, phenomenology, gluon, color, mixing, standard model, electroweak, collider, nucleon, effective theory, sensitivity, new physic, high energy, hadron, chiral perturbation theory, next-to-leading order, impact, neutrino mass, resonance, signal, process, accuracy, collaboration, distribution, flavor, decay, effective field theory, determination, violation, evolution, account, meson, element, baryon, higgs, contribution, gamma & algebra, manifold, geometry, spacetime, partition, modulus space, gravity, theory, branes, correspondence, central charge, deformation, action, chern-simons, duality, string, horizon, supergravity, ad, quantum, space-time, yang-mills, coordinate, entropy, conformal field theory, sitter, field, construction, surface, dimension, boundary, transformation, black hole, solution, mechanic, space, conjecture, type, class, quantization, dirac, formulation, background, connection, massless \\
\end{longtable}

%% file: Table3.tex
\begin{longtable}{p{7cm}|p{7cm}}
\caption{Vocabulary specific to phenomenology (left column) versus experiment (right column).}
\label{table:specific_pheno_vocabulary_exp_ph}\\
\toprule
                                                                                                                                                                                                                                                                                                                                                                                                                                                                             Vocabulary specific to phenomenology &                                                                                                                                                                                                                                                                                                                                                                                                                                              Vocabulary specific to experiment \\
\midrule
\endfirsthead
\caption[]{Vocabulary specific to phenomenology (left column) versus experiment (right column).} \\
\toprule
                                                                                                                                                                                                                                                                                                                                                                                                                                                                             Vocabulary specific to phenomenology &                                                                                                                                                                                                                                                                                                                                                                                                                                              Vocabulary specific to experiment \\
\midrule
\endhead
\midrule
\multicolumn{2}{r}{{Continued on next page}} \\
\midrule
\endfoot

\bottomrule
\endlastfoot
experimental data, qcd, mssm, quark, dark matter, lhc, color, phenomenology, gluon, plasma, new physic, heavy ion collision, account, inflation, parton, evolution, high energy, factorization, effect, implication, scenario, potential, approach, contribution, electroweak, process, mixing, model, estimate, numerical result, accuracy, integral, approximation, neutrino, unification, higgs, possibility, bound, neutrino mass, calculation, case, early universe, sensitivity, generator, extra dimension & detector, sample, measurement, search, upper limit, confidence, experiment, atlas, target, cm, luminosity, event, proton-proton collision, evidence, resolution, fraction, result, gev, first time, beam, expectation, yield, tev, world, top quark, branching, range, technique, muon, limit, study, construction, data, reaction, recent result, mev, system, investigation, section, paper, observation, respect, differential cross section, electron, experimental result \\
\end{longtable}

%% file: Table4.tex
\begin{longtable}[H]{p{0.2\textwidth}|p{0.8\textwidth}}
\caption{Most frequent terms for each topic.}
\label{table:top_words}\\ \midrule
\toprule
{} &                                                                                                                                                                                  Most frequent expressions \\ \midrule
Topic (context)                                       &                                                                                                                                                                                                            \\ \midrule
\midrule
\endfirsthead
\caption[]{Most frequent terms for each topic.} \\ \midrule
\toprule
{} &                                                                                                                                                                                  Most frequent expressions \\ \midrule
Topic (context)                                       &                                                                                                                                                                                                            \\ \midrule
\midrule
\endhead
\midrule
\multicolumn{2}{r}{{Continued on next page}} \\ \midrule
\midrule
\endfoot

\bottomrule
\endlastfoot
Algorithms and calculation techniques                 &                                                               simulation, carlo, monte, lattice, method, correlation, distribution, cluster, generator, statistical, study, function, scaling, size, event \\ \midrule
Amplitude of processes in colliders                   &                                                     amplitude, contribution, state, interaction, resonance, final, final state, process, exchange, reaction, tree, scattering, double, polarization, level \\ \midrule
Amplitudes and Feynman Diagram                        &                                                             amplitude, function, loop, limit, pole, conformal, relation, integral, diagram, correlation, scattering, analytic, block, correlators, feynman \\ \midrule
Analyses and measurements from colliders              &                                                                         data, measurement, event, result, detector, experiment, gev, algorithm, analysis, muon, experimental, energy, precision, fit, beam \\ \midrule
Annihilation and scattering cross-sections            &                                             section, cross, annihilation, photon, energy, scattering, gev, production, total, elastic, process, pair, total cross section, total cross, elastic scattering \\ \midrule
Astrophysics                                          &                                              star, wave, nuclear, matter, neutron, collision, gravitational waves, energy, nuclear matter, flow, density, gravitational, relativistic, heavy-ion, equation \\ \midrule
Black holes                                           &                                black, hole, black hole, black holes, horizon, entropy, extremal, radiation, schwarzschild, thermodynamics, black hole solutions, black hole entropy, hawking, charge, kerr \\
\end{longtable}

%% file: Table5.tex
\begin{longtable}[H]{p{0.25\textwidth}|p{0.6\textwidth}|p{0.15\textwidth}}
\caption{PACS categories most correlated to the topics derived with the unsupervised model. Correlation is measured as the mutual pointwise information (pmi).}
\label{table:full_topics_pacs_pmi}\\
\toprule
                                        &                    &   pmi \\
topic & PACS category &       \\
\midrule
\endfirsthead
\caption[]{PACS categories most correlated to the topics derived with the unsupervised model. Correlation is measured as the mutual pointwise information (pmi).} \\
\toprule
                                        &                    &   pmi \\
topic & PACS category &       \\
\midrule
\endhead
\midrule
\multicolumn{3}{r}{{Continued on next page}} \\
\midrule
\endfoot

\bottomrule
\endlastfoot
\multirow{5}{*}{\begin{tabular}{l}Algorithms and\\ calculation\\ techniques\end{tabular}} & Lattice theory and statistics &  1.39 \\
                                        & Lattice gauge theory &  1.17 \\
                                        & Lattice QCD calculations &  1.12 \\
                                        & Particle correlations and fluctuations &  0.99 \\
                                        & Inelastic scattering: many-particle final states &  0.80 \\
\cline{1-3}
\multirow{5}{*}{\begin{tabular}{l}Amplitude of\\ processes in\\ colliders\end{tabular}} & Baryon resonances (S=C=B=0) &  1.13 \\
                                        & Pion-baryon interactions &  1.10 \\
                                        & Meson-meson interactions &  1.03 \\
\end{longtable}